\documentstyle[preprint,tighten,eqsecnum,aps,prd]{revtex}
\begin{document}
\draft
\preprint{UCSBTH-95-17
\quad hep-th/9507107
\quad
submitted to {\it Physical Review D}}
\title{Generalized quantum mechanics\\
of nonabelian gauge theories}
\author{John T.~Whelan\thanks{Electronic Mail Address:
whelan@cosmic.physics.ucsb.edu}}
\address{Department of Physics, University of California,
Santa Barbara, California 93106-9530}
\date{August 31, 1995}
\maketitle
\begin{abstract}
Hartle's generalized quantum mechanics in the sum-over-histories
formalism is used to describe a nonabelian gauge theory.  Predictions
are made for certain alternatives, with particular attention given to
coarse grainings involving the constraint.  In this way, the theory is
compared to other quantum-mechanical descriptions of gauge theories in
which the constraints are imposed by hand.  The vanishing of the
momentum space constraint is seen to hold, both through a simple
formal argument and via a more careful description of the Lorentzian
path integral as defined on a spacetime lattice.  The configuration
space realization of the constraint is shown to behave in a more
complicated fashion.  For some coarse grainings, we recover the known
result from an abelian theory, that coarse grainings by values of the
constraint either predict its vanishing or fail to decohere.  However,
sets of alternatives defined in terms of a more complicated quantity
in the abelian case are exhibited where definite predictions can be
made which disagree with the assumption that the constraints vanish.
Finally, the configuration space sum-over-histories theory is
exhibited in a manifestly Lorentz-invariant formulation.
\end{abstract}
\narrowtext
\section{Introduction}
General relativity (GR) possesses a symmetry, namely diffeomorphism
invariance.  In the $3+1$ formulation, this divides into time
reparametrization invariance and the nonabelian gauge group of spatial
diffeomorphisms \cite{ADM}.  A technique for formulating quantum
gravity, such as a generalized quantum mechanics defined by a sum over
histories, will have to address the issues raised by these
invariances, such as how (or whether) to enforce the constraints which
the invariances imply.  It is thus useful to examine proposed quantum
formulations of GR by considering simpler theories exhibiting a subset
of these invariances or similar ones.  Previous work has applied the
generalized quantum mechanics program to the
reparametrization-invariant theory of a relativistic world line
\cite{leshouches,relpart} and to abelian gauge theory
\cite{leshouches}.  In this paper, we formulate a generalized quantum
mechanics of a nonabelian gauge theory, and examine the predictions
for some alternatives.

    The role of this work with regard to the vast body of knowledge on
Yang-Mills or nonabelian gauge theories (see \cite{YM} for a review)
is twofold: First, this is the first application of a ``decoherence
functional'' or ``consistent histories'' method to their
quantization.\footnote{Recent work\protect\cite{QCYM} considers
decoherence effects in the quantum cosmology of {\em massive} gauge
fields, However, that work differs from the present enterprise in that
the gauge fields are there seen as a model of matter coupled to
gravity, while we consider massless gauge fields as a toy model for
vacuum gravity itself.  Even more significant is that while they study
decoherence {\em effects}, it is in the context of a WKB quantization
scheme, rather than a generalized quantum mechanics or consistent
histories approach.}  As such, the focus is not primarily upon using
such a theory for the practical consideration of the strong or weak
interaction, but as a toy model which exhibits some features of GR.
However, even as a quantization of a nonabelian gauge theory itself,
both the generalized quantum mechanics formalism and this
implementation thereof deal with different aspects of the theory than
are usually considered.  The alternatives for which generalized
quantum mechanics predicts probabilities are not limited to
projections onto eigenstates of operators at a single moment of time,
but include alternatives defined by field averages over space{\em
time} regions, which are inaccessible in a theory based on states and
wave function reduction.  This broader class of alternatives is
especially of interest in connection with GR, where it is undesirable
to single out a particular time variable for a conventional
quantization. In addition, the considerations herein are predominantly
nonperturbative, as contrasted with the usual perturbative scattering
problems addressed in most practical treatments of Yang-Mills theory.
On the other hand, the rich subject of topological aspects of
nonabelian gauge theories is not considered, and any potential global
properties are in fact ignored by our assumptions about the behavior
of fields at spatial infinity.

    A second accomplishment of this paper is that technical aspects of
the path integrals involved in quantizing a nonabelian gauge theory
are more carefully considered than in the standard literature.
Delicate issues involved in the time slicing of an explicit
(``skeletonized'') construction of the Lorentzian path integral
(sections~\ref{sec:imple} and~\ref{ssec:pired}) are dealt with which
are described only formally or implicitly in standard treatments such
as \cite{fadeev}.  Also, section~\ref{sec:lorentz} exhibits a formal
description of this quantization scheme which is manifestly Lorentz
invariant.

    The plan of this paper is as follows: The review of nonabelian
gauge theory in Section~\ref{sec:formu} establishes the perspective
and notational conventions for the rest of the paper.  We also provide
therein a brief description of the generalized quantum mechanics
formalism and a heuristic recipe for applying it to the theory of
interest.

Section~\ref{sec:imple} describes the explicit implementation of that
application, both as a formal path integral and in a spacetime lattice
approximation\footnote{This is not in the sense of lattice gauge
theory with its Euclidean lattice, Wilson loops, {\it etc.,} but
simply a means to provide what Hatfield\cite{hatfield} calls a
constructive definition of the path integral.} to the path integral.
Technology is developed therein for handling the lattice expressions
(in particular the time slicing) explicitly, which should be of use in
other treatments of path integrals as well.  In the latter half of the
section, we verify explicitly that the implementation is gauge
invariant.  We also show there that our sum-over-histories expression
agrees, in its description of the propagator, with the results of a
reduced phase space canonical operator theory in which the constraints
(Gauss's law) are enforced before quantization.  In our generalized
quantum mechanics formulation, the constraints are not enforced
identically--as they are in a reduced phase space implementation--but
are quantities whose values must be predicted by the theory.  Thus the
prediction of probabilities for the values of the constraints occupies
most of our attention in the remainder of the paper.

     Section~\ref{sec:phase} considers one subset of all possible
alternatives which defines a ``phase space'' realization (as defined
in section~\ref{ssec:phaseallowed}) of the physical gauge fields.  The
predictions of such a theory are found to be consistent with the
vanishing of the constraints for nearly all such sets of alternatives,
and thus to agree with those of a reduced phase space canonical
theory.  In section~\ref{sec:config} we consider another subset of the
allowed alternatives in which the gauge electric field is realized in
terms of the potentials rather than their conjugate momenta.  Since
the momenta are then not restricted by the alternatives, we perform
the integrals over them and reduce our theory to a ``configuration
space'' one.  Now, defining the constraints by their configuration
space realizations, we find the two most significant results of the
paper.  In section~\ref{SSEC:CONFIGK} we see that for some quantities
which vanish when the constraints are satisfied we recover the result
of \cite{leshouches} for electromagnetism, namely, either the
quantities vanish with probability one or quantum mechanical
interference prevents us from assigning probabilities to possible
outcomes.  We also verify that there are coarse grainings which fall
into the first category.  However, the result does not necessarily
hold for all quantities which vanish in the presence of the
constraints, and in section~\ref{SSEC:EMK} we exhibit such a quantity
in the abelian theory of electromagnetism for which we predict a
nonzero probability of an alternative inconsistent with the
constraints.  Since this set of alternatives involves averages of
fields over time, it is not accessible in less general quantum
Yang-Mills theories.  From a spacetime point of view, it is sensible
that the constraints do not have a special status in this theory,
being just one component of the equations of motion.

 Finally, section~\ref{sec:lorentz} verifies that the configuration
space theory is Lorentz-invariant by casting the formal path integral
in a form where that invariance is manifest, even in the attachment of
the initial and final states.

\section{Formulation} \label{sec:formu}
\subsection{Nonabelian gauge theories}
\subsubsection{Fields}
 In this section we set out the conventions used herein to describe a
nonabelian gauge theory (NAGT) in flat spacetime with the metric
$\mathop{\rm diag}\nolimits(-1,1,1,1)$.

The gauge group is described by Hermitian generators $\{T_a\}$ with
real, totally antisymmetric structure constants $\{f_{ab}^c\}$:
$[T_a,T_b]=if_{ab}^c T_c$.\footnote{Repeated indices are
summed over.}

A gauge transformation is described by a matrix $U=e^{ig\Lambda_a(x)
T_a}$.  The connection is a four-vector $A^a(x)$ with components
$\{A^a_\mu(x)\}$, which transforms under infinitesimal gauge
transformations according to
\begin{equation}\label{trans4A}
\delta A^a_\mu = -\nabla\!_\mu\,\delta\Lambda_a
- g f_{ab}^c A^c_\mu \delta\Lambda^b.
\end{equation}
If we define a covariant derivative
\begin{equation}
D_\mu=\nabla\!_\mu + i g A^a_\mu T_a,
\end{equation}
it transforms according to
\begin{equation}
D_\mu\rightarrow UD_\mu U^{-1}.
\end{equation}
This means that if $\psi$ is an isovector, i.e., a vector in the
same space as the matrices $\{T_a\}$ which transforms under gauge
transformations according to $\psi\rightarrow U\psi$, the covariant
gradient of $\psi$ will transform the same way:
\begin{equation}
D_\mu\psi(x)\rightarrow U D_\mu \psi(x).
\end{equation}

The field strength tensor is
\begin{equation}
G^a_{\mu\nu} T_a = {[D_\mu,D_\nu]\over ig}.
\end{equation}

In a particular Lorentz frame, we divide the connection $A^a$ into
scalar and vector potentials $\varphi^a$ and ${\bf A}^a$:
\begin{mathletters}
\begin{eqnarray}
\varphi_a&=&A^0_a\\
{\bf A}_a&=&A^i_a
{\bf e}_i
\end{eqnarray}
\end{mathletters}
and the field strength tensor into gauge electric and magnetic fields
$\bf E$ and $\bf B$:
\begin{mathletters}
\begin{eqnarray}
E^i_a&=&G^{0i}_a\\
B^i_a&=&{1\over 2}\epsilon^{ijk}G^{jk}_a
\end{eqnarray}
\end{mathletters}
where $\epsilon^{ijk}$ is the Levi-Civita symbol.  The gauge electric
and magnetic fields can then be expressed in terms of the scalar and
vector potentials as
\begin{mathletters}
\begin{eqnarray}
{\bf E}_a &=& -\dot{\bf A}_a -\bbox\nabla\varphi_a
- g f_{ab}^c {\bf A}_c \varphi_b \label{defE}\\
{\bf B}_a &=& \bbox\nabla \times {\bf A}_a
+ {1\over 2}g f_{ab}^c {\bf A}_c\times{\bf A}_b\label{defB}.
\end{eqnarray}
\end{mathletters}

The gauge electric and magnetic fields can be shown to transform under
gauge transformations as follows:
\begin{mathletters}
\begin{eqnarray}
{\bf E}_a T_a &\rightarrow& U {\bf E}_a T_a U^{-1}\\
{\bf B}_a T_a &\rightarrow& U {\bf B}_a T_a U^{-1},
\end{eqnarray}
\end{mathletters}
which becomes, for an infinitesimal transformation,
\begin{mathletters}
\begin{eqnarray}
\delta{\bf E}_a &=& - g f_{ab}^c \delta\Lambda_b {\bf E}_c\\
\delta{\bf B}_a &=& - g f_{ab}^c \delta\Lambda_b {\bf B}_c.
\end{eqnarray}
\end{mathletters}
This is the transformation property of an isovector in the adjoint
representation, in which the generators are represented by
$(T^c)_{ab}=-i f_{ab}^c,$
so we will often drop the index from $\bf E$ or $\bf B$ and consider
it to be an isovector in the adjoint representation.  The connection
$A^a$ has an inhomogeneous piece in its transformation law
(\ref{trans4A}), so it is not a true isovector, but we will represent
it as one notationally.  Thus the gauge electric field can be written
as
\begin{equation}\label{adjE}
{\bf E}=-\dot{\bf A}-{\bf D}\varphi,
\end{equation}
where we have realized $\bf D$ in the adjoint representation as
\begin{equation}
{\bf D}_{ab} = \delta_{ab}\bbox\nabla + g f_{ab}^c {\bf A}_c.
\end{equation}

\subsubsection{Classical equations of motion}
The action for a NAGT in the absence of matter is
\begin{equation}\label{configact}
S=\int d^4\!x {\cal L}=-\int d^4\!x {1\over 4} G_a^{\mu\nu}G^a_{\mu\nu}
=\int d^4\!x {1\over 2}({\bf E}^2 -{\bf B}^2).
\end{equation}
The conjugate momenta in a particular reference frame are found by
differentiating the Lagrangian density $\cal L$ with respect to
$\dot A^\mu=\partial_t A^\mu$:
\begin{mathletters}
\begin{eqnarray}
\pi_0&=&{{\cal D}{\cal L}\over{\cal D}\dot\varphi}=0\\
\bbox\pi&=&{{\cal D}{\cal L}\over{\cal D}\dot{\bf A}}
=\dot{\bf A} + {\bf D}\varphi=-{\bf E}.\label{conj}
\end{eqnarray}
\end{mathletters}
The Hamiltonian density is given by
\begin{eqnarray}
{\cal H}[A,\bbox\pi]&=&\dot{\bf A}\cdot\bbox\pi-{\cal L}
={1\over 2}\bbox\pi^2+{1\over 2}{\bf B}^2-\bbox\pi\cdot{\bf D}\varphi
\nonumber\\
&=&{\cal H}[{\bf A},\bbox\pi]-\bbox\pi\cdot{\bf D}\varphi\label{ham}
\end{eqnarray}
and Hamilton's equations of motion are
\begin{mathletters}
\begin{eqnarray}
\dot{\bf A}&=&\bbox\pi - {\bf D}\varphi\label{piE}\\
D_t \bbox\pi &=& {\bf D\times B}\label{curlB}\\
{\bf D\cdot\bbox\pi}&=&0.\label{constraint}
\end{eqnarray}
\end{mathletters}
Equation (\ref{constraint}) involves no time derivatives, so it is the
constraint of this NAGT, which we call $K={\bf D}\cdot\bbox\pi$.

\subsection{Generalized quantum mechanics}
\subsubsection{Formalism and definitions}\label{sssec:GQM}
     We will use Hartle's generalized quantum mechanics
formalism\cite{gqm}.  The three fundamental elements of this theory
are the possible histories of the system (``fine-grained histories''),
allowable partitions of the histories into classes $\{c_\alpha\}$ so
that each history is contained in exactly one class (``coarse
grainings''), and a complex matrix $D(\alpha,\alpha ')$ corresponding
to each coarse graining (``decoherence functional'').  The decoherence
functional must satisfy the following properties.
\begin{mathletters}

 Hermiticity:
          \begin{equation}\label{herm}
          D(\alpha ',\alpha)=D^*(\alpha,\alpha ').
          \end{equation}

 Positivity of diagonal elements:
          \begin{equation}\label{pos}
          D(\alpha,\alpha)\ge 0.
          \end{equation}

 Normalization:
          \begin{equation}
          \sum_\alpha\sum_{\alpha '} D(\alpha,\alpha ')=1.
          \end{equation}

 Superposition: If $\{c_\beta\}$ is a coarse graining
constructed by combining classes in $\{c_\alpha\}$ to form larger
classes (``a coarser graining''), i.e., $c_\beta =
\bigcup\limits_{\alpha\in\beta} c_\alpha$, the decoherence functional
for $\{c_\beta\}$ can be constructed from the one for $\{c_\alpha\}$
by
          \begin{equation}\label{super}
          D(\beta,\beta')=\sum_{\alpha\in\beta}\sum_{\alpha '\in\beta'}
D(\alpha,\alpha ').
          \end{equation}
\end{mathletters}

     When the decoherence functional is diagonal, or nearly so:
\begin{equation}\label{medium}
D(\alpha,{\alpha '})\approx\delta_{\alpha{\alpha '}}p_\alpha,
\end{equation}
we say the alternatives exhibit medium decoherence, and identify the
diagonal elements $\{p_\alpha\}$ as probabilities of the alternatives
$\{c_\alpha\}$.  In fact, all that is necessary for the $\{p_\alpha\}$
to obey the probability sum rules is a less restrictive condition
known as {\em weak} decoherence:
\begin{equation}\label{weak}
\mathop{\rm Re}\nolimits D(\alpha,{\alpha '})
\approx\delta_{\alpha{\alpha '}}p_\alpha.
\end{equation}
When we refer to ``decoherence'' with no modifier, we will implicitly
mean medium decoherence.  When the alternatives do not decohere at
least weakly, quantum mechanical interference prevents the theory from
assigning probabilities to them.

    As an example of how a decoherence functional is constructed,
consider nonrelativistic operator quantum mechanics.  The decoherence
functional is a generalization of the usual formula for the
probability that a measurement at a time $t$ of a quantity modeled by
an operator $\widehat Q$ will yield a result in a range
$\Delta_\alpha$, if the system is described by a density matrix $\rho$
at time $t'$:
\begin{eqnarray}
p_\alpha&=&\int\limits_{Q\in\Delta_\alpha}dQ\langle Q|
e^{-i(t-t')\widehat H}\rho e^{i(t-t')\widehat H}|Q\rangle
\nonumber\\
&=&\mathop{\rm Tr}\nolimits
\left[P_\alpha e^{-i(t-t')\widehat H}\rho e^{i(t-t')\widehat H}\right].
\label{prob}
\end{eqnarray}
A possible outcome of the measurement (an alternative) is described
by a projection operator
\begin{equation}
P_\alpha=\int\limits_{Q\in\Delta_\alpha}dQ|Q\rangle\langle Q|,
\end{equation}
and the complete set of alternatives obeys
\begin{mathletters}
\begin{eqnarray}
\sum_\alpha P_\alpha&=&\openone
\\
P_\alpha P_\beta&=&\delta_{\alpha\beta}P_\alpha.
\end{eqnarray}
\end{mathletters}
If we generalize to the case of an initial state $\rho'$ and a final
state $\rho''$ (the case of ``no final condition'' is described by
$\rho''=\openone$), as well as a series of $n$ sets of alternatives at
times $\{t_i\}$, each of which is described by a set of projection
operators $\{P^i_{\alpha_i}\}$, where
\begin{mathletters}
\begin{eqnarray}\label{projsum}
\sum_{\alpha_i} P_{\alpha_i}&=&\openone
\\
P_{\alpha_i} P_{\beta_i}&=&\delta_{\alpha_i\beta_i}P_{\alpha_i},
\end{eqnarray}
\end{mathletters}
the description in terms of probabilities must be replaced by a
decoherence functional description:
\begin{equation}\label{decop}
D(\alpha,\alpha')
={\mathop{\rm Tr}\nolimits[\rho''C_\alpha\rho'C_{\alpha'}^\dagger]
\over
\mathop{\rm Tr}\nolimits
[\rho''e^{-i(t''-t')\widehat H}\rho' e^{i(t''-t')\widehat H}]},
\end{equation}
 where $C_\alpha$ is the chain of projections corresponding to a
particular series of alternatives out of the series of sets:
\begin{eqnarray}
C_\alpha&=&e^{-i(t''-t_n)\widehat{H}}P^n_{\alpha_n}
e^{-i(t_n-t_{n-1})\widehat{H}}P^{n-1}_{\alpha_{n-1}}\ldots
\nonumber\\
&&\times\ldots
P^2_{\alpha_2}e^{-i(t_2-t_1)\widehat{H}}P^1_{\alpha_1}
e^{-i(t_1-t')\widehat{H}}.
\label{chain}
\end{eqnarray}
 It is straightforward to show that when the class operator consists
of a single projection ($C_\alpha=e^{-i(t''-t)\widehat{H}}P_\alpha
e^{-i(t-t')\widehat{H}}$) and there is no final condition
($\rho''=\openone$), the decoherence functional reduces exactly to the
diagonal form in (\ref{medium}) with the probabilities given by the
familiar form (\ref{prob}).

As a consequence of (\ref{projsum}) the class operator
obeys
\begin{equation}
\sum_\alpha C_\alpha=e^{-i(t''-t')\widehat{H}}.
\end{equation}

    In the sum-over-histories formulation (which exists for the
preceding nonrelativistic quantum mechanics example as well as for the
NAGT which is the focus of this paper), the chain of projections
$C_\alpha$ is replaced by a class operator whose matrix elements are
defined via a path integral, as described below.

\subsubsection{Application to a NAGT}
To formulate a NAGT in generalized quantum mechanics, we follow a
procedure similar to the one described in \cite{leshouches} for
electromagnetism (E\&M).

    Since we want to express the theory in terms of a sum over
histories without reference to a Hilbert space, we replace the initial
density matrix $\rho'$ with a set of wave functionals\footnote{We
establish the convention here that the arguments of functionals are
enclosed in square brackets, and that a field with a prime or index
such as $A''$ or $A^M$ indicates a field configuration as a function
of spatial position $\bf x$ only, while an unadorned field such as $A$
refers to a function defined for all spacetime points $x$.}
$\{\Psi_j[A']\}$ with corresponding non-negative weights (or
``probabilities'') $\{p'_j\}$.  (In a Hilbert space theory this would
mean defining $\rho'=\sum_j|\Psi_j\rangle p'_j\langle\Psi_j|$.)
Similarly, the final state is now defined by a set of wave functionals
$\{\Phi_i[A'']\}$ and weights $\{p''_i\}$, which replace the density
matrix $\rho''$.

The wave functionals are taken to be functionals of scalar and vector
potential configurations on an initial or final surface of constant
time, as appropriate.\footnote{$\Psi[A',t')$ is a functional of field
configurations \{$A'({\bf x)}$\} and a function of time $t'$, whence
the hybrid parentheses.} They are assumed to obey the operator form of
the constraints $\pi_0=0$ and ${\bf D}\cdot\bbox\pi=0$:
\begin{mathletters}\label{Psicons}
\begin{eqnarray}
{{\cal D}\over{\cal D}\varphi'}\Psi[A',t')&=&0
\label{phiind}\\
{\bf D}\cdot{{\cal D}\over{\cal D}{\bf A'}}\Psi[A',t')&=&0
\label{Psisym}
\end{eqnarray}
\end{mathletters}
 and likewise for $\Phi[A'',t'')$.  (Since the wave functionals are
independent of the scalar potential $\varphi$, we will henceforth
write the first argument as the three-vector ${\bf A}'$ rather than
the four-vector $A'$.)  Equation (\ref{Psisym}) is analogous to the
momentum constraint in GR, while (\ref{phiind}) corresponds to the
lack of dependence of the wave functional for quantum GR on the shift
vector $N_i$.

With these conventions, we replace the definition (\ref{decop}) for
the decoherence functional with
\begin{equation}\label{dec}
D(\alpha,{\alpha '})
={\sum\limits_{i,j}p_i''\langle\Phi_i|C_\alpha|\Psi_j\rangle
\langle\Phi_i|C_{\alpha '}|\Psi_j\rangle^* p_j'
\over
\sum\limits_{i,j}p_i''|\langle\Phi_i|C_u|\Psi_j\rangle|^2 p_j'}.
\end{equation}
 Here the quantity $\langle\Phi_i|C_\alpha|\Psi_j\rangle$ is analogous
to a matrix element of the class operator for the class $c_\alpha$,
but it is constructed by a sum over the histories in the class
$c_\alpha$, weighted by the initial and final wave functionals $\Psi_j$
and $\Phi_i$ evaluated at the endpoints ${\bf A}'$ and ${\bf A}''$ of
the history.  Schematically:
\begin{equation}\label{SOH}
\langle\Phi_i|C_\alpha|\Psi_j\rangle
=\sum_{\text{history}\in\alpha}
\Phi_i^*[{\bf A}'']e^{iS[\text{history}]}\Psi_j[{\bf A}'].
\end{equation}
 It is convenient to refer to a ``class operator $C_\alpha$'' even in
the sum-over-histories theory, and when we do, we mean the object
defined by (\ref{SOH}).  $C_u$ is the class operator corresponding to
the class $c_u$ of all paths, which is just the propagator.

Having described schematically the construction of the decoherence
functional, we now specify the other two elements which describe the
generalized quantum mechanics.  The fine grained histories summed over
are complete field configurations $A(x)$ [and also $\bbox\pi(x)$ if we
are considering a phase space formulation] in the region between the
initial and final time slices.  The allowable coarse grainings are
limited to gauge invariant partitions of the fields.

\section{Class operators in the path integral formulation}
\label{sec:imple}
\subsection{Overview}
    This section describes in detail how to implement the sum over
histories heuristically described in (\ref{SOH}).  In
section~\ref{ssec:formal} we express this as a formal path integral.
Section~\ref{ssec:lattice} contains an explicit realization of this
integral on a discrete spacetime lattice, where the lattice spacing is
to be taken to be infinitesimally small.\footnote{It should be stated
once again that we are {\em not} doing Lattice Gauge Theory in
anything like the usual sense.  The action is expressed directly in
terms of fields defined at each lattice point and not in terms of the
``links'' defining a ``plaquette''.  In addition, our spacetime
lattice is Lorentzian rather than Euclidean.} The following two
sections demonstrate that the particular details chosen in
section~\ref{ssec:lattice} were suitable by showing that the path
integral has desired properties. In section~\ref{ssec:lattred}, the
sum-over-histories expression for the class operator $C_u$
corresponding to the class $c_u$ of all paths is shown to equal, up to
a constant multiplicative factor, the propagator $e^{i\widehat{H}_{\rm
red}T}$ in a reduced phase space canonical theory. In
section~\ref{ssec:invar} the path integral is shown to be unchanged
under the discrete equivalent of a gauge transformation, in the limit
that the lattice spacing goes to zero.

\subsection{Formal expression}\label{ssec:formal}
     We express the sum over histories (\ref{SOH}) via a path integral
which is formally written as
\begin{eqnarray}\label{formalphase}
&&\langle\Phi|C_\alpha|\Psi\rangle=
\\
&&\int\limits_\alpha{\cal D}^4\! A{\cal D}^3\pi
\Phi^*\!\left[{\bf A}'',t''\right)\delta[G]\Delta_G[A,\bbox\pi]
e^{iS_{\rm can}[A,\bbox\pi]}\Psi[{\bf A}',t'),
\nonumber
\end{eqnarray}
 where the gauge condition is $G=0$ and $\Delta_G$ is the
corresponding Fadeev-Popov gauge-fixing determinant.  It was
originally defined in \cite{fadeev} in terms of the Poisson bracket:
\begin{mathletters}
\begin{equation}
\Delta_G=|\det\{G,K\}|.\label{fppb}
\end{equation}
 Other useful (and equivalent) definitions are (see e.g.,
\cite{hatfield})
\begin{equation}\label{fpdet}
\Delta_G
=\left|
     \det\left[{{\cal D}G^{\Lambda}\over{\cal D}\Lambda}\right]
\right|
\end{equation}
 and
\begin{equation}\label{fpint}
{1\over\Delta_G}=\int{\cal D}\Lambda\,\delta[G^{\Lambda}-G],
\end{equation}
\end{mathletters}
 where $\Lambda$ is the parameter defining a gauge transformation
which takes $G$ into $G^{\Lambda}$.

 Finally, the canonical action is
\begin{eqnarray}
S_{\rm can}&=&\int d^4\!x
\left(
	\dot{\bf A}\cdot\bbox\pi - {\cal H}[A,\bbox\pi]
\right)
\nonumber\\
&=&\int d^4\!x \left(\dot{\bf A}\cdot\bbox\pi - {1\over 2}\bbox\pi^2
- {1\over 2}{\bf B}^2 + \bbox\pi\cdot{\bf D}\varphi\right).
\label{canact}
\end{eqnarray}

If we assume\footnote{Throughout this paper we will neglect any global
issues such as the Gribov ambiguity\cite{gribov} and assume that
fields can be taken to vanish at spatial infinity.} that $\varphi$
vanishes at spatial infinity, we can integrate by parts\footnote{It is
worth pointing out once explicitly that the covariant gradient $\bf D$
behaves like the ordinary gradient under integration by parts.
Examining $\alpha{\bf D}\beta
=\alpha_a\bbox\nabla\beta_a+f_{ab}^c\alpha_a\beta_b{\bf A}_c$, we see
that the first term integrates by parts as usual, and the second term
also picks up a minus sign under the interchange of $\alpha$ and
$\beta$ due to the antisymmetry of $f_{ab}^c$.  Thus $\alpha{\bf
D}\beta =\bbox\nabla(\alpha\beta)-\beta{\bf D}\alpha$.} to obtain
\begin{equation}
S_{\rm can}=\int d^4\!x
\left(
     \dot{\bf A}\cdot\bbox\pi - {1\over 2}\bbox\pi^2
     - {1\over 2}{\bf B}^2 - \varphi{\bf D}\cdot\bbox\pi
\right).
\end{equation}

    The expression (\ref{formalphase}) involves the full set of phase
space variables, but we will also use it as the starting point for the
configuration space formulation.  If our coarse graining makes no
reference to the conjugate momentum $\bbox\pi$, we can work in a gauge
which does not restrict $\bbox\pi$ and integrate it out to obtain
\begin{equation}\label{formconf}
\langle\Phi|C_\alpha|\Psi\rangle
=\int\limits_\alpha{\cal D}^4\! A\,\Phi^*[{\bf A}'',t'')
\delta[G]\Delta_G[A]e^{iS[A]}\Psi[{\bf A}',t'),
\end{equation}
 where $S$ is the (configuration space) action (\ref{configact}) and
for the purposes of this formal expression, a constant factor has been
absorbed into ${\cal D}^4\!A$.

\subsection{Lattice realization}
\label{ssec:lattice}
     To give a concrete meaning to the formal path integral in
(\ref{formalphase}), we imagine it to be defined on an arbitrarily
small lattice; the spatial volume is divided into lattice elements of
volume $\delta^3\!x$ and the time interval from $t'$ to $t''$ is
divided into slices of separation $\delta t={t''-t' \over J+1}\equiv{T
\over J+1}$.  The lattice expression for the class operator is then
\widetext
\begin{eqnarray}
\langle\Phi|C_\alpha|\Psi\rangle
&=& \int{\cal D}^4\! A^{J+1}\Phi^*\!\left[{\bf A}^{J+1},t''\right)
\nonumber\\
&&\times
\Bigglb(
     \prod_{M=J}^0\int{\cal D}^4\! A^M{\cal D}^3\pi^M\exp
     \left\{i\delta t\int d^3\!x
          \left(
               \dot{\bf A}^M\cdot\bbox\pi^M
               -{\cal H}\left[\overline{\bf A}^M\!,\bbox\pi^M\right]
               -\overline{\varphi}^M\overline{\bf D}^M\!\cdot\bbox\pi^M
          \right)
     \right\}
\Biggrb)
\nonumber\\
&&\times
\left(\prod_{M=0}^{J+2}\delta\left[G^M\right]\Delta_{G^M}\right)
\Psi\left[{\bf A}^0,t'\right)e_\alpha\left[A,\bbox\pi\right],
\label{latticephase}
\end{eqnarray}
\narrowtext
where the ``barred'' quantities indicate temporal averages:
\addtocounter{equation}{-1}
\begin{mathletters}
\begin{eqnarray}
\overline{A}^M&=&{A^{M+1}+A^M \over 2},\quad 0\le M\le J\\
\overline{\bf D}^M_{ab}
&=&\delta_{ab}\bbox\nabla + g f_{ab}^c\overline{\bf A}^M_c
\end{eqnarray}
and the lattice expression for the ``velocity'' is
\begin{equation}\label{velo}
\dot{\bf A}^M={{\bf A}^{M+1}-{\bf A}^M \over \delta t},
\quad 0\le M\le J.
\end{equation}
\end{mathletters}
 The expression (\ref{latticephase}) reflects the fact that since the
``velocity'' $\dot{\bf A}^M$ is naturally associated with a point {\em
halfway} between the coordinate lattice slices labeled $M$ and $M+1$
by (\ref{velo}), it is sensible to associate the conjugate momenta
$\bbox\pi^M$ with those points as well, in light of the term $\dot{\bf
A}^M\cdot\bbox\pi^M$.  This means that to maintain manifest time
reversal symmetry we should not associate $\bbox\pi^M$ with $A^M$ or
$A^{M+1}$, but instead with $\overline{A}^M$.  The gauge-fixing
expressions $G^M$ and $\Delta_{G^M}$ are also assumed to be expressed
in terms of the averaged fields $\left\{\overline{A}^M\right\}$
whenever their complexity prevents an unambiguous definition in terms
of the $\left\{A^M\right\}$ alone.\footnote{The reason why there are
$J+3$ gauge conditions $\left\{G^M\right\}$ is most easily seen in the
temporal gauge $\varphi\equiv0$.  On a lattice this corresponds to the
$J+2$ conditions $\left\{\varphi^M=0, 0\le M\le J+1\right\}$.
However, there is residual gauge freedom in the temporal gauge, since
a gauge transformation by a parameter $\Lambda({\bf x})$ [see
(\ref{trans4A})] which is independent of the time $t$ will preserve
the temporal gauge condition $\varphi\equiv 0$.  To completely fix the
gauge, then, we would need to specify one other quantity over all
space at a particular time.  This is the last of the $J+3$ gauge
conditions.}

The factor of $e_\alpha$ is a functional of the paths which is unity
for any path in the class $c_\alpha$ and vanishes for any path not in
$c_\alpha$.

The remaining functional integrals $\left\{{\cal D}^4\!A^M\right\}$
and $\left\{{\cal D}^3\pi^M\right\}$ are over functions of the
spatial co\"{o}rdinate $\bf x$.  We leave consideration of the spatial
dependence somewhat formal, because all of the complications involved
in describing the gradients on the lattice basically appear, and more
seriously, in the temporal direction.  We will thus be speaking as
though the spacetime function $A(x)$ is broken up into a series of
functions of a continuous spatial variable $\bf x$: $\left\{A^M({\bf
x})\right\}$.  However, we can ultimately consider the following
lattice resolutions for the functional integrals and delta functions:
\begin{mathletters}\label{meas}
\begin{eqnarray}
{\cal D}A^{M\mu}&=&\prod_{a,\bf x}N_A d A^{M\mu}_a({\bf x})\\
{\cal D}\pi^{Mi}&=&\prod_{a,\bf x}N_\pi d\pi^{Mi}_a({\bf x})\\
\delta\left[A^{M\mu}\right]&=&\prod_{a,\bf x}
\delta\biglb(A^{M\mu}_a({\bf x})\bigrb)/N_A\\
\delta\left[\pi^{Mi}\right]&=&\prod_{a,\bf x}
\delta\biglb(\pi^{Mi}_a({\bf x})\bigrb)/N_\pi,
\end{eqnarray}
\end{mathletters}
where $N_A$ and $N_\pi$ are arbitrary normalization constants which
obey
\begin{equation}\label{NANp}
N_A N_\pi={\delta^3\!x\over 2\pi}.
\end{equation}
 \{This definition is chosen to give the desirable properties
(\ref{eigennorm}) for potential and momentum eigenstates in the
corresponding operator theory.  See also equation~(6.228) of
\cite{swanson}.\}

    Since (\ref{NANp}) only defines a relation between $N_A$ and
$N_\pi$, there is still an arbitrary factor in the definitions of the
measures (\ref{meas}).  It is reassuring to verify that the class
operator matrix elements $\langle\Phi|C_\alpha|\Psi\rangle$ defined by
(\ref{latticephase}) are independent of that arbitrary factor so that
for instance if we double $N_A$ (and thus halve $N_\pi$), they are
unchanged.  To do this correctly, we must also keep in mind that the
wave functionals $\Psi[{\bf A}']$ and $\Phi[{\bf A}'']$ also depend on
the value of $N_A$ as follows.  Let the wave functionals be normalized
according to
\begin{equation}\label{wavenorm}
\int{\cal D}^3\!A'\delta\bbox[G[{\bf A}']\bbox]\Delta_G[{\bf A}']
\bigl|\Psi\left[{\bf A}'\right]\bigr|^2=1.
\end{equation}
Since there is a factor of ${\cal N}_A^3$ (where ${\cal
N}_A=\prod_{a,\bf x}N_A$ and ${\cal N}_\pi=\prod_{a,\bf x}N_\pi$)
associated with the measure ${\cal D}^3\!A'$ and a factor of ${\cal
N}_A^{-1}$ associated with the gauge fixing delta function
$\delta[G']$, there must be a factor of ${\cal N}_A^{-2}$ associated
with the square of the wave functional so that the factors all cancel
out.  That is, if we double ${\cal N}_A$, we must halve $\Psi[{\bf
A}']$ to maintain the normalization (\ref{wavenorm}).  Considering the
expression (\ref{latticephase}), if we multiply together all the
normalization factors (${\cal N}_A$ for each ${\cal D}A^{M\mu}$,
${\cal N}_\pi$ for each ${\cal D}\pi^{Mi}$, ${\cal N}_A^{-1}$ for each
$\delta[G_M]$, and ${\cal N}_A^{-1}$ for each wave functional) we have
\begin{eqnarray}
&&{\cal N}_A^4 {\cal N}_A^{-1} ({\cal N}_A^4 {\cal N}_\pi^3)^{J+1}
({\cal N}_A^{-1})^{J+3} {\cal N}_A^{-1}
\nonumber\\
&&={\cal N}_A^{3J+3} {\cal N}_\pi^{3J+3}
=\prod_{a,\bf x}\left({\delta^3\!x\over 2\pi}\right)^{3J+3}
\end{eqnarray}
and the arbitrariness of the normalization $N_A$ does indeed cancel
out of (\ref{latticephase}).

\subsection{Relation to reduced phase space theory}\label{ssec:lattred}

 To illustrate why the details of (\ref{latticephase}) were chosen, we
show in this section that it gives the same expression, up to a
normalization constant, for the propagator $C_u$ as would be obtained
from a canonical theory working with only the ``physical degrees of
freedom''.  [Comparisons of (\ref{latticephase}) to the reduced phase
space results for particular coarse grainings will be considered in
later sections.] This derivation is essentially the one given in
\cite{fadeev}, taken in reverse order and with more attention paid to
the details of the lattice.

    First, we observe that there are $J+2$ co\"{o}rdinate variables
$\left\{A^M({\bf x})\right\}$ but only $J+1$ averaged variables
$\left\{\overline{A}^M({\bf x})\right\}$.  We can define
$\overline{A}^{J+1}$ to be a linear combination of the
$\left\{A^M\right\}$ independent of the $J+1$ other
$\left\{\overline{A}^M\right\}$, normalized so that the change of
variables from $\left\{A^M\right\}$ to $\left\{\overline{A}^M\right\}$
has unit determinant and
\begin{equation}
\prod_{M=0}^{J+1} dA^{M\mu}_a({\bf x})
= \prod_{M=0}^{J+1} d\overline{A}^{M\mu}_a({\bf x}).
\end{equation}
\{One such choice is
$\overline{A}^{J+1}=2^J[A^{J+1}+(-1)^{J+1}A^0]$.\} Since neither the
velocity term $\dot{\bf A}\cdot\bbox\pi$ nor the initial and final
wave functionals depends on the scalar potential $\varphi$, it only
enters (\ref{latticephase}) through the ``genuine'' averages
$\left\{\overline{\varphi}^M, 0\le M\le J\right\}$, and thus the class
operator is independent of $\overline{\varphi}^{J+1}$.  It is then
natural to choose $\overline{\varphi}^{J+1}({\bf x})=0$ as one of our
$J+3$ time slices worth of gauge conditions.  The corresponding
Fadeev-Popov determinant is
\begin{equation}\label{fpphi}
\Delta_{\overline{\varphi}^{J+1}}=\det[D_t]|_{\varphi=0}
=\det[\partial_t],
\end{equation}
which is a constant.  Overall constant (i.e., the same for all
$\alpha$) factors in the class operator
$\langle\Phi|C_\alpha|\Psi\rangle$ will cancel out in the expression
(\ref{dec}) for the decoherence functional and will not effect the
physics.

    For this demonstration, it is simplest to choose as the bulk of
the gauge conditions the axial gauge, in which the component of
$\bf A$ along some fixed unit vector ${\bf e}_n$ vanishes:
\begin{equation}\label{ax}
{\bf e}_n\cdot{\bf A}(x)\equiv A_n(x)=0.
\end{equation}
The Fadeev-Popov determinant of this gauge condition is
\begin{equation}
\Delta_{A_n}=\det[-D_n]|_{A_n=0}=\det[-\partial_n],
\end{equation}
another constant.  The remaining components of the vector potential
are
\begin{equation}
{\bf A}_\perp={\bf A}-A_n{\bf e}_n.
\end{equation}

    Thus the class operator becomes
\widetext
\ifpreprintsty
\begin{eqnarray}
&&\langle\Phi|C_\alpha|\Psi\rangle=
\int{\cal D}^2\!A_\perp^{J+1}
\Phi^*\!\left[{\bf A}_\perp^{J+1},t''\right)
\det[\partial_t]\det[-\partial_n]\nonumber\\
&&\times
\Bigglb(
	\prod_{M=J}^0\int{\cal D}^2\!A_\perp^M
        {\cal D}\overline{\varphi}^M{\cal D}^3\pi^M
	\exp\left\{
		i\delta t\int d^3\!x
		\left(\dot{\bf A}_\perp^M\cdot\bbox\pi_\perp^M
		-{\cal H}
		\left[\overline{\bf A}_\perp^M,\bbox\pi^M\right]
		-\overline{\varphi}^M
		\overline{\bf D}^M\!\cdot\bbox\pi^M\right)
	\right\}
	\det[-\partial_n]
\Biggrb)\nonumber\\
&&\times\Psi\left[{\bf A}_\perp^0,t'\right)e_\alpha[A,\bbox\pi].
\label{classax}
\end{eqnarray}
\else
\begin{eqnarray}
\langle\Phi|C_\alpha|\Psi\rangle&=&
\int{\cal D}^2\!A_\perp^{J+1}
\Phi^*\!\left[{\bf A}_\perp^{J+1},t''\right)
\det[\partial_t]\det[-\partial_n]\nonumber\\
&&\times
\Bigglb(
	\prod_{M=J}^0\int{\cal D}^2\!A_\perp^M
        {\cal D}\overline{\varphi}^M{\cal D}^3\pi^M
	\exp\left\{
		i\delta t\int d^3\!x
		\left(\dot{\bf A}_\perp^M\cdot\bbox\pi_\perp^M
		-{\cal H}
		\left[\overline{\bf A}_\perp^M,\bbox\pi^M\right]
		-\overline{\varphi}^M
		\overline{\bf D}^M\!\cdot\bbox\pi^M\right)
	\right\}
	\det[-\partial_n]
\Biggrb)\nonumber\\
&&\times\Psi\left[{\bf A}_\perp^0,t'\right)e_\alpha[A,\bbox\pi].
\label{classax}
\end{eqnarray}
\fi

    To specialize to the propagator $C_u$, which is defined by a sum
over {\em all} paths, we set $e_\alpha=1$.  We can perform each of the
$\overline{\varphi}$ integrals to obtain
\begin{eqnarray}
\int{\cal D}\overline{\varphi}^M\exp
\left(
     -i\delta t\int d^3\!x\,\overline{\varphi}^M
     \overline{\bf D}^M\!\cdot\bbox\pi^M
\right)
&=&\prod_{a,\bf x}\int N_A d\overline{\varphi}^M_a({\bf x})\exp
\left[
     -i\delta t\,\delta^3\!x\,\overline{\varphi}_a^M({\bf x})
     \left(
          \overline{\bf D}^M\!\cdot\bbox\pi^M
     \right)
     _a({\bf x})
\right]
\nonumber\\
&=&\prod_{a,\bf x}N_A{2\pi\over\delta^3\!x\delta t}\delta
\Biglb(
     \left(
          \overline{\bf D}^M\!\cdot\bbox\pi^M
     \right)
     _a({\bf x})
\Bigrb)
=\delta
\left[
     \overline{\bf D}^M\!\cdot\bbox\pi^M
\right]
\delta T^{-1},\label{phiint}
\end{eqnarray}
\narrowtext
where we have defined the infinite constant
\begin{equation}
\delta T=\prod_{a,\bf x}\delta t.
\end{equation}

    In a reduced phase space theory, the gauge component (here $A_n$)
of the co\"{o}rdinate is taken to vanish, and the corresponding
component of the conjugate momentum is restricted to the value which
causes the constraint to be satisfied.  [Since the Lagrange
multipliers (here the scalar potential $\varphi$) multiply identically
enforced constraints, they do not appear in the Lagrangian.]  In the
axial gauge, this means that $\pi_n$ has the value which ensures
$K={\bf D}\cdot\bbox\pi=0$, or
\begin{equation}
D_n\pi_n=\partial_n\pi_n=-{\bf D}_\perp\cdot\bbox\pi_\perp.
\end{equation}
We define a functional which accomplishes that:
\begin{eqnarray}
\pi_n\left[{\bf A}'_\perp,\bbox\pi'_\perp,{\bf x}\right)
&=&\int^{x_n} dx'_n
(-{\bf D}'_\perp\cdot\bbox\pi'_\perp)({\bf x}_\perp+{\bf e}_nx'_n)
\nonumber\\
&=&-\left(\partial^{-1}_n {\bf D}'_\perp\cdot\bbox\pi'_\perp\right)
({\bf x}).
\label{pin}
\end{eqnarray}

	Returning to the sum-over-histories expression, we observe
that the delta function from (\ref{phiint}) enforcing the constraint
can be rewritten
\begin{equation}\label{deltaK}
\delta\left[\overline{\bf D}^M\!\cdot\bbox\pi^M\right]
=\delta\Biglb[\pi_n^M
-\pi_n\left[\overline{\bf A}^M_\perp,\bbox\pi^M_\perp\right]\Bigrb]
\left|\det\left[{{\cal D}\pi^M_n\over{\cal D}K^M}\right]\right|.
\end{equation}
 Recalling (\ref{fppb}) we identify the determinant in the expression
above as the reciprocal of the Fadeev-Popov determinant, since
\begin{equation}
\Delta_{A_n}=\bigl|\det[\{A_n,K\}]\bigr|
=\det\left[{{\cal D}K\over{\cal D}\pi_n}\right]
=\det[D_n].
\end{equation}
Thus we can combine the Fadeev-Popov determinant with the $\varphi$
integral producing the delta function (\ref{deltaK}) to obtain
\begin{eqnarray}
&&\delta\left[\overline{A}_n^M\right]\Delta_{\overline{A}_n^M}
\int{\cal D}\overline{\varphi}^M\exp
\left(
     -i\delta t\int d^3\!x\,\overline{\varphi}^M
     \overline{\bf D}^M\!\cdot\bbox\pi^M
\right)
\nonumber\\
&&=\delta\left[\overline{A}_n^M\right]
\delta\Biglb[\pi_n^M
-\pi_n\left[\overline{\bf A}^M_\perp,\bbox\pi^M_\perp\right]\Bigrb]
\delta T^{-1},
\end{eqnarray}
which means we can rewrite the propagator as
\begin{eqnarray}
&&\langle\Phi|C_u|\Psi\rangle=
\int{\cal D}^2\!A_\perp^{J+1}
\Phi^*\!\left[{\bf A}_\perp^{J+1},t''\right)
\det[\partial_t]\det[-\partial_n]
\nonumber\\
&&\times\delta T^{-1}
\Bigglb(
    \prod_{M=J}^0\int{\cal D}^2\! A_\perp^M{\cal D}^2\pi_\perp^M
\nonumber\\
    &&\qquad\times\exp
    \left\{
          i\delta t\int d^3\!x
          \left(
               \dot{\bf A}_\perp^M\cdot\bbox\pi_\perp^M
               -{\cal H}_{\rm red}
               \left[
                    \overline{\bf A}^M_\perp,\bbox\pi^M_\perp
               \right]
          \right)
    \right\}
\Biggrb)
\nonumber\\
&&\times\Psi\left[{\bf A}_\perp^0,t'\right),
\label{redpath}
\end{eqnarray}
where ${\cal H}_{\rm red}$ is the reduced Hamiltonian density
\begin{equation}\label{Hred}
{\cal H}_{\rm red}[{\bf A}'_\perp,\bbox\pi'_\perp]
={\cal H}
\bbox[
     {\bf A}'_\perp,\bbox\pi'_\perp+{\bf e}_n\pi_n
     [
            {\bf A}'_\perp,\bbox\pi'_\perp
     ]
\bbox].
\end{equation}

    To convert (\ref{redpath}) into a canonical operator form of the
propagator, we consider eigenstates of the ``physical''
co\"{o}ordinate and momentum operators:
\begin{equation}
\widehat{\bf A}_\perp|{\bf A}_\perp'\rangle
={\bf A}_\perp'|{\bf A}_\perp'\rangle\qquad
\widehat{\bbox\pi}_\perp|\bbox\pi_\perp'\rangle
=\bbox\pi_\perp'|\bbox\pi_\perp'\rangle,
\end{equation}
normalized so that
\begin{mathletters}\label{eigennorm}
\begin{eqnarray}
\langle{\bf A}_\perp''|{\bf A}_\perp'\rangle
&=&\delta[{\bf A}_\perp''-{\bf A}_\perp']\qquad
\\
\langle{\bf A}_\perp'|\bbox\pi_\perp'\rangle
&=&\exp\left(i\int d^3\!x\,{\bf A}_\perp'\cdot\bbox\pi_\perp'\right),
\end{eqnarray}
which implies, via (\ref{meas}-\ref{NANp}), that
\begin{equation}
\langle\bbox\pi_\perp''|\bbox\pi_\perp'\rangle
=\delta[\bbox\pi_\perp''-\bbox\pi_\perp'].
\end{equation}
\end{mathletters}
A state vector corresponding to each wave functional is then defined
by
\begin{equation}
|\Psi\rangle=\int{\cal D}^2\!A'_\perp
|{\bf A}'_\perp\rangle\Psi[{\bf A}'_\perp].
\end{equation}

    The operator form of the reduced Hamiltonian density has an
ambiguity because of the operator ordering of the term ${1\over
2}\left(\pi_n[{\bf A}'_\perp,\bbox\pi'_\perp]\right)^2$.  A
natural\footnote{\protect{\label{fn:opord}}Different choices of
operator ordering will typically lead to different lattice
realizations than the one in (\ref{latticephase}), although no
systematic description of the correspondence is known to the author.
Two simple examples of alternate operator ordering are one in which
all the $\widehat{\bf A}$'s are placed to the right of all the
$\widehat{\bbox\pi}$'s and one in which the $\widehat{\bbox\pi}$'s are
to the right of the $\widehat{\bf A}$'s.  The former will lead to a
lattice expression in which $\bbox\pi^M$ and $\dot{\bf A}^M$ are
associated with $A^M$, while the latter will associate $\bbox\pi^M$
and $\dot{\bf A}^M$ with $A^{M+1}$.  Weyl ordering is preferable to
either of these two because the midpoint rule to which it leads does
not pick out the future or the past as a preferred direction in time.
The midpoint rule was also originally advocated by Feynman
\{equation~(20) in \cite{feyn}\} as the natural skeletonization of a
path.} choice is Weyl ordering\cite{weyl}, which has the property that
\widetext
\begin{eqnarray}
\langle{\bf A}_\perp''|\widehat{\cal W}
\left(
     F\left[{\bf A}_\perp,\bbox\pi_\perp\right]
\right)
|{\bf A}_\perp'\rangle
&=&\int{\cal D}^2\pi'_\perp F
\left[
     {{\bf A}'_\perp+{\bf A}''_\perp\over 2},\bbox\pi'_\perp
\right]
\langle{\bf A}''_\perp|\bbox\pi'_\perp\rangle
\langle\bbox\pi'_\perp|{\bf A}'_\perp\rangle\nonumber\\
&=&\int{\cal D}^2\pi'_\perp F
\left[
     {{\bf A}'_\perp+{\bf A}''_\perp\over 2},\bbox\pi'_\perp
\right]
\exp
\left[
     i\int d^3\!x ({\bf A}''_\perp-{\bf A}'_\perp)\cdot\bbox\pi'_\perp
\right].
\end{eqnarray}
Thus, defining $\widehat{\cal H}_{\rm red}
=\widehat{\cal{W}}({\cal{H}}_{\rm red})$, we can rewrite
\begin{eqnarray}
&&\int{\cal D}^2\pi_\perp^M\exp
\left\{
     i\delta t\int d^3\!x
     \left(
          \dot{\bf A}_\perp^M\cdot\bbox\pi_\perp^M
          -{\cal H}_{\rm red}
          \left[
               \overline{\bf A}^M_\perp,\bbox\pi^M_\perp
          \right]
     \right)
\right\}\nonumber\\
&&\qquad=\int{\cal D}^2\pi_\perp^M
\left\{
     1-i\delta t\int d^3\!x{\cal H}_{\rm red}
     \left[
          {{\bf A}^M_\perp+{\bf A}^{M+1}_\perp\over 2},\bbox\pi^M_\perp
     \right]
     +O\biglb((\delta t)^2\bigrb)
\right\}
\exp
\left[
     i\int d^3\!x
     \left(
          {\bf A}_\perp^{M+1}-{\bf A}_\perp^M
     \right)
     \cdot\bbox\pi_\perp^M
\right]\nonumber\\
&&\qquad=\left<{\bf A}^{M+1}_\perp\right|
\left[
     1-i\delta t\int d^3\!x\widehat{\cal H}_{\rm red}
     +O\biglb((\delta t)^2\bigrb)
\right]
\left|{\bf A}^M_\perp\right>,
\label{incprop}
\end{eqnarray}
so that, dropping terms of order $(\delta t)^2$,
\begin{eqnarray}
\langle\Phi|C_u|\Psi\rangle&=&
\int{\cal D}^2\!A_\perp^{J+1}
\left<\Phi(t'')\left|{\bf A}_\perp^{J+1}\right.\right>
\det[\partial_t]\det[-\partial_n]\delta T^{-1}\nonumber\\
&&\times\left[
     \prod_{M=J}^0\int{\cal D}^2\! A_\perp^M
     \left<{\bf A}^{M+1}_\perp\right|\exp
     \left(-i\delta t\int d^3\!x\widehat{\cal H}_{\rm red}\right)
     \left|{\bf A}^M_\perp\right>
\right]
\left<\left.{\bf A}_\perp^0\right|\Psi(t')\right>.
\end{eqnarray}
\narrowtext
Using the fact that
\begin{equation}
\int{\cal D}^2\!A_\perp^M|{\bf A}^M_\perp\rangle\langle{\bf A}^M_\perp|
=1,
\end{equation}
we see that this differs from the operator expression
\begin{equation}
\langle\Phi|e^{-i\widehat{H}_{\rm red} T}|\Psi\rangle
\end{equation}
only by the factor of $\det[\partial_t]\det[-\partial_n]\delta
T^{-1}$, which is a constant.  {From} (\ref{dec}), we see that
multiplying the class operator by a constant factor has no effect on
the decoherence functional.

\subsection{Gauge invariance}\label{ssec:invar}

We will now show that the theory described by (\ref{latticephase}) is
gauge invariant.  We do this explicitly and in detail because the
standard demonstration \cite{fadeev} makes use of a canonical
transformation, which should be ill-defined at the endpoints of the
integration due to the fact that the path integral
(\ref{latticephase}) has one more configuration space integration than
phase space integration.  Put otherwise, if $\overline{\bf A}^M$ and
$\bbox\pi^M$ are linked by a canonical transformation, there is no
$\bbox\pi^{J+1}$ to which the extra degree of freedom $\overline{\bf
A}^{J+1}$ corresponds.

First, we must describe how to implement an
infinitesimal gauge transformation
\begin{mathletters}\label{contgauge}
\begin{eqnarray}
\delta A^a_\mu&=&-\nabla_\mu\delta\Lambda^a
- g f_{ab}^c A^c_\mu\delta\Lambda^b
\\
\delta\bbox\pi^a &=& - g f_{ab}^c \bbox\pi^c\delta\Lambda^b
\end{eqnarray}
\end{mathletters}
on a lattice.  We replace the continuous function $\delta\Lambda(x)$
with functions $\left\{\delta\Lambda^M({\bf x})\right\}$ defined on
lattice slices 0 through $J+1$.  The transformation for the vector
potential can then be taken as
\begin{equation}\label{gaugelatt}
\delta{\bf A}_a^M=-\bbox\nabla\delta\Lambda_a^M
- g f_{ab}^c {\bf A}_c^M\delta\Lambda_b^M,
\end{equation}
but there is no simple translation of the scalar potential
transformation law because of the time derivative.  More practical
than the transformation of ${\bf A}^M$ given in (\ref{gaugelatt}) will
be its implications for transformations of the time derivative
$\dot{\bf A}^M$ and midpoint average $\overline{\bf A}^M$.  Using the
fact that
\begin{eqnarray}
&&{A^{M+1}B^{M+1}-A^M B^M \over \delta t}
\nonumber\\
&&\quad={\left(A^{M+1}-A^M\right)\over\delta t}
{\left(B^{M+1}+B^M\right)\over 2}
\nonumber \\
&&\quad\quad+{\left(A^{M+1}+A^M\right)\over 2}
{\left(B^{M+1}-B^M\right)\over\delta t}
\nonumber\\
&&\quad=\dot{A}^M\overline{B}^M+\overline{A}^M\dot{B}^M,
\end{eqnarray}
we see that (\ref{gaugelatt}) implies
\begin{equation}\label{dAdot}
\delta\dot{\bf A}_a^M=-\bbox\nabla\delta\dot\Lambda_a^M
- g f_{ab}^c \left(\dot{\bf A}_c^M\overline{\delta\Lambda}_b^M
+\overline{\bf A}_c^M\delta\dot\Lambda_b^M\right),
\end{equation}
which is exactly what one might write down from the corresponding
continuum expression
\begin{equation}
\delta\dot{\bf A}_a=-\bbox\nabla\delta\dot\Lambda_a
- g f_{ab}^c \left(\dot{\bf A}_c\delta\Lambda_b
+{\bf A}_c\delta\dot\Lambda_b\right).
\end{equation}
The case is not quite so simple with the averages.  Since
\begin{eqnarray}
&&{1\over 2}\left(A^{M+1}B^{M+1}+A^M B^M\right)
\nonumber\\
&&\quad={1\over 4}\left(A^{M+1}+A^M\right)\left(B^{M+1}+B^M\right)
\nonumber\\
&&\quad\quad+{1\over 4}\left(A^{M+1}-A^M\right)\left(B^{M+1}-B^M\right)
\nonumber\\
&&\quad=\overline{A}^M\overline{B}^M
+{(\delta t)^2\over 4}\dot{A}^M\dot{B}^M,
\end{eqnarray}
we have
\begin{equation}\label{dAbar}
\delta\overline{\bf A}_a^M=-\bbox\nabla\overline{\delta\Lambda}_a^M
-g f_{ab}^c \left(\overline{\bf A}_c^M\overline{\delta\Lambda}_b^M
+{(\delta t)^2\over 4}\dot{\bf A}_c^M\delta\dot{\Lambda}_b^M\right),
\end{equation}
which differs from the na\"{\i}ve analog of (\ref{gaugelatt}) by a
term proportional to $(\delta t)^2$.  However, we can neglect this
term by the following familiar argument \cite{FnH}: as the lattice
spacing $\delta t$ goes to zero, the factor of $\exp[-i\delta t\int
d^3\!x (\bbox\pi^2/2-\bbox\pi\cdot\dot{\bf A})]$ will oscillate
rapidly and suppress the path integral if $\dot{\bf A}$ is more
singular than $(\delta t)^{-1/2}$.  Likewise, if we concern ourselves
only with gauge transformations which take histories which are
sufficiently nonsingular to contribute to the path integral into other
such histories, (\ref{dAdot}) tells us that the $\delta\dot{\Lambda}$
should also blow up no faster than $(\delta t)^{-1/2}$ as $\delta
t\rightarrow 0$.  Counting the factors of $\delta t$ in the upper
bounds, we see that the extra term in (\ref{dAbar}) should be at worst
proportional to $\delta t$ and thus be negligible for sufficiently
small lattice spacing.  Thus we can use the simpler formula
\begin{equation}\eqnum{\protect{\ref{dAbar}}$'$}
\delta\overline{\bf A}_a^M=-\bbox\nabla\overline{\delta\Lambda}_a^M
-g f_{ab}^c \overline{\bf A}_c^M\overline{\delta\Lambda}_b^M.
\end{equation}

The transformation laws for the conjugate momentum and scalar
potential can be defined by analogy to (\ref{dAbar}$'$).  As observed
in Sec.~\ref{ssec:lattred} the scalar potential only enters the class
operator via its averaged values (assuming that
$\overline{\varphi}^{J+1}=0$ is always taken as a gauge choice), and
so we only need to know how to transform $\overline{\varphi}$ and not
$\varphi$.  Since $\overline{\varphi}$ and $\bbox\pi$ are defined
midway between lattice points, we prescribe transformation laws which
are the obvious lattice realizations of (\ref{contgauge}):
\begin{mathletters}
\begin{eqnarray}
\delta\overline{\varphi}_a^M&=&\delta\dot{\Lambda}_a^M
-g f_{ab}^c \overline{\varphi}_c^M\overline{\delta\Lambda}_b^M\\
\delta\bbox\pi_a^M&=&-g f_{ab}^c
\bbox\pi_c^M\overline{\delta\Lambda}_b^M.\label{dpi}
\end{eqnarray}
\end{mathletters}

Now we show that under such a gauge transformation, the expression
(\ref{latticephase}) for the class operator is unchanged.  First, we
examine the measure for the path integral.  The demonstrations for
${\cal D}^3\!A^M$, ${\cal D}\overline{\varphi}^M$ and ${\cal
D}^3\!\pi^M$ are all essentially the same, so we show it explicitly
only for ${\cal D}^3\!A^M$.  Under the gauge transformation ${\bf
A}^M\rightarrow \widetilde{\bf A}^M={\bf A}^M+\delta{\bf A}^M$, we
have
\begin{equation}
\prod_a d\widetilde{A}_a^{Mi}({\bf x})
=\left(\prod_a dA_a^{Mi}({\bf x})\right)
\det\left\{{\partial\widetilde{A}_b^{Mi}({\bf x})\over
\partial A_c^{Mi}({\bf x})}\right\}.
\end{equation}
The Jacobian matrix is
\begin{equation}
{\partial\widetilde{A}_b^{Mi}({\bf x})
\over\partial A_c^{Mi}({\bf x})}
=\delta_{bc}+
{\partial\,\delta A_b^{Mi}({\bf x})\over \partial A_c^{Mi}({\bf x})}
=\delta_{bc}-g f_{bd}^c\delta\Lambda_d^M({\bf x}).
\end{equation}
Using the standard matrix result that to lowest order in $\delta a$,
$\det(1+\delta a)=1+\mathop{\rm Tr}\nolimits\delta a$, we see that the
Jacobian for the transformation is
\begin{equation}
\det\left\{{\partial\widetilde{A}_b^{Mi}({\bf x})\over
\partial A_c^{Mi}({\bf x})}\right\}
=1-f_{bd}^b\delta\Lambda_d^M({\bf x})=1
\end{equation}
where we have used the fact that the structure constants are totally
antisymmetric.  This tells us that
\begin{mathletters}\label{meastrans}
\begin{equation}
{\cal D}^3\!\widetilde{A}^M={\cal D}^3\!A^M,
\end{equation}
and the demonstrations that
\begin{equation}
{\cal D}\overline{\widetilde{\varphi}}{}^M
={\cal D}\overline{\varphi}^M
\end{equation}
and
\begin{equation}
{\cal D}^3\widetilde{\pi}^M={\cal D}^3\pi^M
\end{equation}
\end{mathletters}
proceed similarly.

    Next we consider the canonical action density [i.e., $\dot{\bf
A}\cdot\bbox\pi$ minus the Hamiltonian density; cf.{\
}(\ref{canact})].  The demonstration is simplest if we undo the
integration by parts to write it in the form
\begin{eqnarray}
&&\dot{\bf A}^M\cdot\bbox\pi^M
+\bbox\pi^M\cdot\overline{\bf D}^M\overline{\varphi}^M
-{1\over 2}\left(\bbox\pi^M\right)^2-{1\over 2}\left({\bf B}^M\right)^2
\nonumber\\
&&\qquad=
-{\bf E}^M\cdot\bbox\pi^M-{1\over 2}\left(\bbox\pi^M\right)^2
-{1\over 2}\left({\bf B}^M\right)^2,
\label{undo}
\end{eqnarray}
 where we have defined the lattice realizations of the electric and
magnetic fields by analogy to (\ref{adjE}) and (\ref{defB}):
\begin{mathletters} \label{EB}
\begin{equation}
{\bf E}^M=-\dot{\bf A}^M-\overline{\bf D}^M\overline{\varphi}^M
\end{equation}
and
\begin{equation}
{\bf B}_a^M=\bbox\nabla\times\overline{\bf A}_a^M
+{1\over 2}gf_{ab}^c\overline{\bf A}_c^M\times\overline{\bf A}_b^M.
\end{equation}
\end{mathletters}
The conjugate momentum $\bbox\pi$ is defined by (\ref{dpi}) to
transform as an isovector.  The transformations of ${\bf E}$ and ${\bf
B}$ are
\begin{mathletters}
\begin{eqnarray}
\delta{\bf E}_a^M&=&
-\delta\dot{\bf A}_a^M
-\left({\bf D}\delta\overline{\varphi}^M\right)_a-g f_{ab}^c
\delta\overline{\bf A}_c^M \overline{\varphi}_b^M\\
\delta{\bf B}_a^M
&=&\bbox\nabla\times\delta\overline{\bf A}_a^M +g f_{ab}^c
\delta\overline{\bf A}_c^M\times\overline{\bf A}_b^M,
\end{eqnarray}
\end{mathletters}
which can be shown, with a little algebra, to give the expected
isovector transformations:
\begin{mathletters}
\begin{eqnarray}
\delta{\bf E}_a^M&=&
-g f_{ab}^c {\bf E}_c^M\overline{\delta\Lambda}_b^M
\\
\delta{\bf B}_a^M&=&
-g f_{ab}^c {\bf B}_c^M\overline{\delta\Lambda}_b^M.
\end{eqnarray}
\end{mathletters}
 So, since $\bbox\pi^M$, ${\bf E}^M$ and ${\bf B}^M$ all transform as
isovectors, $\left(\bbox\pi^M\right)^2$, $\left({\bf B}^M\right)^2$ and
${\bf E}^M\cdot\bbox\pi^M$ are gauge invariant quantities, and
\begin{eqnarray}
&&\dot{\widetilde{\bf A}}{}^M\cdot\widetilde{\bbox\pi}^M
+\widetilde{\bbox\pi}^M \cdot
\overline{\widetilde{\bf D}}{}^M\overline{\widetilde{\varphi}}^M
-{1\over 2}\left(\widetilde{\bbox\pi}^M\right)^2
-{1\over 2}\left(\widetilde{\bf B}^M\right)^2
\nonumber\\
&&=\dot{\bf A}^M\cdot\bbox\pi^M
+\bbox\pi^M\cdot\overline{\bf D}^M\overline{\varphi}^M
-{1\over 2}\left(\bbox\pi^M\right)^2
-{1\over 2}\left({\bf B}^M\right)^2.
\nonumber\\
\label{acttrans}
\end{eqnarray}

    The gauge fixing delta function becomes
$\delta\left[\widetilde{G}^M\right]$.  Using (\ref{fpint}) to write the
Fadeev-Popov determinant as
\begin{equation}
{1\over\Delta_G}=\int{\cal D}\Lambda'\delta[G^{\Lambda'}-G],
\end{equation}
 where $G^{\Lambda'}$ indicates the result of a gauge transformation
on $G$ by the dummy variable $\Lambda'(x)$, we see that the effect of
a gauge transformation by $\Lambda(x)$ is
\begin{equation}
{1\over\widetilde{\Delta}_G}={1\over\Delta^\Lambda_G}
=\int{\cal D}\Lambda'\delta[G^{\Lambda'\cdot\Lambda}-G^\Lambda].
\end{equation}
 Changing the variable of integration to
$\Lambda''=\Lambda'\cdot\Lambda$ (the gauge transformation
accomplished by successive application of $\Lambda'$ and $\Lambda$),
we have (using formal invariance of the group measure; see
section~7.5 of \cite{swanson} for more details)
\begin{equation}\label{dettrans}
{1\over\widetilde{\Delta}_G}
=\int{\cal D}\Lambda''\delta[G^{\Lambda''}-\widetilde{G}]
={1\over\Delta_{\widetilde{G}}}.
\end{equation}

Coarse graining only by gauge-invariant alternatives means that a path
$\{\widetilde{A},\widetilde{\bbox\pi}\}$ is in the class $c_\alpha$ if
and only if the corresponding path $\{A,\bbox\pi\}$ is, so
\begin{equation}\label{indtrans}
e_\alpha[\widetilde{A},\widetilde{\bbox\pi}]=e_\alpha[A,\bbox\pi].
\end{equation}

    Finally, we consider the behavior of the wave functionals $\Phi^*$
and $\Psi$.
\begin{eqnarray}
\Psi[\widetilde{\bf A}']&=&\Psi[{\bf A}'+\delta{\bf A}']
=\Psi[{\bf A}']
+\int d^3\!x\,\delta{\bf A}'({\bf x})
\cdot{{\cal D}\Psi\over{\cal D}{\bf A}'({\bf x})}
\nonumber\\
&=&\Psi[{\bf A}']
-\int d^3\!x\,({\bf D}'\delta\Lambda')({\bf x})
\cdot{{\cal D}\Psi\over{\cal D}{\bf A}'({\bf x})}.
\end{eqnarray}
Upon integrating by parts, this becomes
\begin{equation}
\Psi[\widetilde{\bf A}']
=\Psi[{\bf A}']
+\int d^3\!x\,\delta\Lambda'\,{\bf D}'\cdot
{{\cal D}\Psi\over{\cal D}{\bf A}'}
=\Psi[{\bf A}']\label{wftrans},
\end{equation}
where we have used (\ref{Psisym}) in the last step.

    Now, we relabel the variables $A$ and $\bbox\pi$ in
(\ref{latticephase}) by $\widetilde{A}$ and $\widetilde{\bbox\pi}$ and
use (\ref{meastrans}), ({\ref{acttrans}), (\ref{dettrans}),
(\ref{indtrans}) and (\ref{wftrans}) to convert the expression to
\widetext
\begin{eqnarray}
\langle\Phi|C_\alpha|\Psi\rangle&=&
\int{\cal D}^3\!\widetilde{A}^{J+1}
\Phi^*\!\left[\widetilde{\bf A}^{J+1},t''\right)\nonumber\\
&&\times
\Bigglb(
     \prod_{M=J}^0\int{\cal D}^3\!\widetilde{A}^M
     {\cal D}\overline{\widetilde{\varphi}}^M
     {\cal D}^3\widetilde{\pi}^M\exp
     \left\{
          i\delta t\int d^3\!x
          \left(
          \dot{\widetilde{\bf A}}^M\!\cdot\widetilde{\bbox\pi}^M
          -{\cal H}
          \left[
               \overline{\widetilde{\bf A}}{}^M\!,
               \widetilde{\bbox\pi}^M
          \right]
          -\overline{\widetilde{\varphi}}^M
          \overline{\widetilde{\bf D}}^M\!\cdot\widetilde{\bbox\pi}^M
          \right)
     \right\}
\Biggrb)\nonumber\\
&&\times
\left(
     \prod_{M=0}^{J+1}\delta
     \left[\widetilde{G}^M\right]\widetilde{\Delta}_{G^M}
\right)
\Psi\left[\widetilde{\bf A}^0,t'\right)
e_\alpha\bigl[\widetilde{A},\widetilde{\bbox\pi}\bigr]\nonumber\\
&=&
\int{\cal D}^3\!A^{J+1}\Phi^*\!\left[{\bf A}^{J+1},t''\right)
\nonumber\\
&&\times
\Bigglb(
     \prod_{M=J}^0\int{\cal D}^3\!A^M{\cal D}\overline{\varphi}^M
     {\cal D}^3\pi^M \exp
     \left\{
          i\delta t\int d^3\!x
          \left(
          \dot{\bf A}^M\!\cdot\bbox\pi^M
          -{\cal H}\left[\overline{\bf A}^M\!,\bbox\pi^M\right]
          -\overline{\varphi}^M\overline{\bf D}^M\!\cdot\bbox\pi^M
          \right)
     \right\}
\Biggrb)\nonumber\\
&&\times
\left(
     \prod_{M=0}^{J+1}\delta
     \left[\widetilde{G}^M\right]\Delta_{\widetilde{G}^M}
\right)
\Psi\left[{\bf A}^0,t'\right)e_\alpha[A,\bbox\pi],
\end{eqnarray}
\narrowtext
which shows that the expressions for the class operators are the same
whether the gauge is $G=0$ or $\widetilde{G}=0$.

\section{Phase space results}\label{sec:phase}

\subsection{Allowed alternatives}\label{ssec:phaseallowed}

    In the previous section, we described a sum-over-histories
construction of the class operators (and hence the decoherence
functional) for a NAGT.  For the class operators to have the desirable
properties detailed therein, the alternatives considered need only be
gauge-invariant coarse grainings of the connection $A$ and the
conjugate momentum $\bbox\pi$.  However, in practice we will not be
interested in arbitrary gauge-invariant quantities, but only the
physical quantities of the theory, namely the gauge electric and
magnetic fields and the covariant derivative.  In a phase space
formulation, those are identified with the isovectors $-\bbox\pi$,
$\bf B$ and $D_\mu$, respectively.  Thus we define physical phase
space coarse grainings to be those in which the gauge electric field
is identified with $-\bbox\pi$ and the gauge magnetic field with $\bf
B$.  (The physical phase space theory is then a subset of the phase
space theory described in section~\ref{sec:imple}.)

We can partition the histories by the values of arbitrary isoscalars
(gauge invariant quantities) constructed from the physically allowed
isovectors.  Since the length of an isovector is an isoscalar, coarse
graining by the length of isovectors is allowed.  It is more
convenient to think of this sort of coarse graining as specifying in
which of a set of regions in isospace an isovector lies, where all the
regions are rotationally invariant.  From now on, when we talk about
coarse graining by isovectors, this is what we mean.

\subsection{Constraints}

So a general ``physical'' coarse graining can involve functionals of
$\bbox\pi$, $\bf B$, $\bf D$, and $D_t$.  If we consider the subset of
coarse grainings which involves the first three but not the covariant
time derivative $D_t$, we see that it involves only the vector
potential $\bf A$ and the conjugate momentum $\bbox\pi$, and not the
scalar potential $\varphi$.  If we work in a gauge which also leaves
$\varphi$ unrestricted, we can perform the path integral over
$\varphi$ to obtain
\begin{equation}
\int{\cal D}\varphi
\exp\left(i\int d^4\!x\,\varphi\,{\bf D}\cdot\bbox\pi\right)
=\delta[{\bf D}\cdot\bbox\pi/2\pi],
\end{equation}
and (\ref{formalphase}) becomes
\begin{eqnarray}\label{fixK}
\langle\Phi|C_\alpha|\Psi\rangle&=&
\int\limits_\alpha{\cal D}^3\!A{\cal D}^3\pi
\Phi^*\!\left[{\bf A}'',t''\right)\det[2\pi]
\\
&&\quad\times\delta[K]\delta[G]\Delta_G[A,\bbox\pi]
e^{iS_{\rm can}[A,\bbox\pi]}\Psi[{\bf A}',t').
\nonumber
\end{eqnarray}

    In particular, if we coarse grain by values of the constraint
$K={\bf D}\cdot\bbox\pi$, the delta function will cause the class
operator to vanish for any class which does not include ${\bf
D}\cdot\bbox\pi=0$, i.e., the constraint satisfied.  If exactly one
class includes that condition, then, it will have the only
non-vanishing class operator, and the only non-vanishing element of
the decoherence functional will be the diagonal element corresponding
to that alternative.  This will then allow the assignment of
probabilities, namely, a probability of 1 for the alternative in which
the constraint is satisfied and 0 for all others.

    Now we make this formal demonstration more precise. Given our
choice of lattice expressions, it should be clear that the relevant
quantities are the following functions defined on each slice:
\begin{equation}
\left\{K^M({\bf x})
=\left(\overline{\bf D}^M\!\cdot\bbox\pi^M\right)({\bf x})\right\}.
\end{equation}
If we work in a gauge which sets $\overline{\varphi}^{J+1}$ to zero
and does not otherwise restrict $\varphi$, we can perform the
$\overline{\varphi}$ integrals as in (\ref{phiint}) and obtain from
(\ref{latticephase}) [recalling (\ref{fpphi})] \widetext
\begin{eqnarray}
\langle\Phi|C_\alpha|\Psi\rangle&=&
\int{\cal D}^3\!A^{J+1}\Phi^*\!\left[{\bf A}^{J+1},t''\right)
\det[\partial_t]
\nonumber\\
&&\times
\Bigglb(
     \prod_{M=J}^0\int{\cal D}^3\!A^M{\cal D}^3\pi^M\exp
     \left\{i\delta t\int d^3\!x
          \left(
               \dot{\bf A}^M\cdot\bbox\pi^M
               -{\cal H}\left[\overline{\bf A}^M\!,\bbox\pi^M\right]
          \right)
     \right\}
\delta\left[K^M\right]\delta T^{-1}
\Biggrb)
\nonumber\\
&&\times\left(\prod_{M=0}^{J+1}\delta\left[G^M\right]
\Delta_{G^M}\right)
\Psi\left[{\bf A}^0,t'\right)e_\alpha\left[A,\bbox\pi\right].
\label{lattfix}
\end{eqnarray}
\narrowtext
 The $\delta\left[K^M\right]$ causes the class operator to vanish for
any coarse graining not consistent with all the $\left\{K^M\right\}$
vanishing everywhere.  Thus if we coarse grain by an average of the
constraint over some spacetime region, which will correspond to some
average over the $\left\{K^M\right\}$, the only non-zero element of
the decoherence functional will be the diagonal one corresponding to
that average vanishing.

\subsection{Comparison to reduced phase space theory}\
\label{ssec:pired}

    In a reduced phase space canonical theory, as described in
section~\ref{ssec:lattred}, the alternatives are defined by
projections onto eigenstates of physical operators.  Working in the
axial gauge, the operator corresponding to the gauge electric field is
[cf.\ (\ref{pin})]
\begin{mathletters}
\begin{equation}
-\left(
    \widehat{\bbox\pi}_\perp
    -{\bf e}_n\partial_n^{-1}
    \widehat{\bf D}_\perp\cdot\widehat{\bbox\pi}_\perp
\right)
\end{equation}
and that corresponding to the gauge magnetic field is
\begin{equation}
\widehat{\bf B}^a=\bbox\nabla\times\widehat{\bf A}^a_\perp
+{1\over 2}gf_{ab}^c
\widehat{\bf A}_\perp^c\times\widehat{\bf A}_\perp^b.
\end{equation}
\end{mathletters}
The operator for the covariant gradient is $\widehat{\bf
D}^{ab}=\delta_{ab}\bbox\nabla+gf_{ab}^c\widehat{\bf A}^c_\perp$, but
it is less clear how to convert
$D^{ab}_t=\delta_{ab}\partial_t-gf_{ab}^c\varphi_c$ into an operator
built out of $\widehat{\bf A}_\perp$ and $\widehat{\bbox\pi}_\perp$.
First of all, we need to consider the time derivatives of
$\widehat{\bf A}_\perp$ and $\widehat{\bbox\pi}_\perp$.  In an
operator theory, one identifies the time derivatives at one moment of
time via the Heisenberg equations of motion:
\begin{mathletters}\label{heis}
\begin{eqnarray}
\partial_t\widehat{\bf A}_\perp
&=&i\left[\widehat{H}_{\rm red},\widehat{\bf A}_\perp\right]
\nonumber\\
&=&\widehat{\bbox\pi}_\perp
+\widehat{\cal W}
\left(
    {\bf D}_\perp\partial_n^{-2}{\bf D}_\perp\cdot\bbox\pi_\perp
\right)
\\
\partial_t\widehat{\bbox\pi}_\perp^a
&=&i\left[\widehat{H}_{\rm red},\widehat{\bbox\pi}_\perp^a\right]
\nonumber\\
&=&-\openone_\perp
\cdot\left(\widehat{\bf D}_\perp\times\widehat{\bf B}\right)^a
\nonumber\\
&&-gf_{ab}^c
\widehat{\cal W}
\Biglb(
    \left(\partial_n^{-2}{\bf D}_\perp\cdot\bbox\pi_\perp\right)^c
    \bbox\pi^b_\perp
\Bigrb),
\end{eqnarray}
\end{mathletters}
 where
\begin{equation}
\openone_\perp=\openone-{\bf e}_n\otimes{\bf e}_n
\end{equation}
is the tensor which projects onto the ``perpendicular'' directions. If
we identify
\begin{equation}
\widehat{\varphi}
=-\partial_n^{-2}\widehat{\bf D}_\perp\cdot\widehat{\bbox\pi}_\perp,
\end{equation}
and remember that the final operator expressions should always be
Weyl-ordered, equations (\ref{heis}) give the operator versions of
$-{\bf E}_\perp=\bbox\pi_\perp$ [two of the components of (\ref{piE})]
and the Maxwell's equation $D_t\bbox\pi_\perp=-({\bf D\times
B})_\perp$ [two of the components of (\ref{curlB})], respectively.  We
also have
\begin{equation}
\widehat{\pi}_n=\partial_n\widehat{\varphi},
\end{equation}
 which is the third component of (\ref{piE}).
So the operator realization of $D_t^{ab}$ is
\begin{equation}
\delta_{ab}\partial_t
+gf_{ab}^c\partial_n^{-2}
\left(\widehat{\bf D}_\perp\cdot\widehat{\bbox\pi}_\perp\right)_c,
\end{equation}
 where the effects of $\partial_t$ on other operators are given by
(\ref{heis}).  Of course, in a sum-over-histories formulation,
attempts to describe instantaneous values of time derivatives do not
yield sensible results due to the non-differentiability of the paths.
Instead, we coarse grain by time derivatives averaged over time, which
correspond to coarse grainings by the difference between values of a
quantity at two finitely separated instants of time.

    As described in section~\ref{sssec:GQM}, probabilities in the
usual operator quantum mechanics are described by expectation values
of projection operators, while in an operator generalized quantum
mechanics \cite{gqm}, an alternative $c_\alpha$ corresponds to a class
operator $C_\alpha$ which is defined as in (\ref{chain}) by a chain of
such projections:
\begin{eqnarray}\label{cop}
C_\alpha&=& e^{-i\widehat{H}_{\rm red} (t''-t_n)} P^n_{\alpha_n}
\\
&&\times
\left(
	\prod_{i=n-1}^1 e^{-i\widehat{H}_{\rm red} (t_{i+1}-t_i)}
	P^i_{\alpha_i}
\right)
e^{-i\widehat{H}_{\rm red}(t_1-t')}.
\nonumber
\end{eqnarray}
 If we generalize further, and allow a class operator to be not just a
single chain of projections, but a sum of such chains, we can describe
more general alternatives, such as coarse grainings by time averages.
The operator expression corresponding to a coarse-grained class
operator would be defined by assigning to each class $c_\alpha$ a sum
of chains of projections:
\begin{eqnarray}\eqnum{\protect{\ref{cop}}$'$}
C_\alpha&=&\sum_{\{\alpha_i\}\in\alpha}
e^{-i\widehat{H}_{\rm red} (t''-t_n)} P^n_{\alpha_n}
\\
&&\times
\left(
	\prod_{i=n-1}^1 e^{-i\widehat{H}_{\rm red}(t_{i+1}-t_i)}
	 P^i_{\alpha_i}
\right)
e^{-i\widehat{H}_{\rm red}(t_1-t')}.
\nonumber
\end{eqnarray}
 This will be equivalent to the corresponding sum-over-histories
expression if we can replace the projections with restricted
integrations on lattice slices.  First, consider a projection onto a
range of the co\"{o}rdinate ${\bf A}_\perp$:
\begin{equation}
P^i_{\Delta_i}=\int {\cal D}^2\!A^\iota_\perp
\left|{\bf A}^\iota_\perp\right>
e_{\Delta_i}[{\bf A}^\iota_\perp]
\left<{\bf A}^\iota_\perp\right|,
\end{equation}
where
\begin{equation}
e_{\Delta_i}[{\bf A}^\iota_\perp]
=\cases{0,&${\bf A}^\iota_\perp\notin\Delta_i$\cr 1,
&${\bf A}^\iota_\perp\in\Delta_i$\cr}
\end{equation}
 is the indicator function for the region $\Delta_i$ in the space of
field configurations $\{{\bf A}^\iota_\perp({\bf x})\}$. To examine
the effect that this projection has on the class operator, assume that
we have taken our lattice spacing small enough that $\delta
t<t_{i+1}-t_i, t_i-t_{i-1}$ so that if $t_I$ is the latest time slice
before $t_i$ (i.e., $t_{I+1}\ge t_i\ge t_I$), no other
projections lie in the interval $(t_I,t_{I+1})$.  Then the effect of
the projection is to modify the right-hand side of (\ref{incprop}) to
be, to first order in $\delta t$,
\begin{eqnarray}
\left<{\bf A}^{I+1}_\perp\right|
&&
\left[1-i(t_{I+1}-t_i)\widehat{H}_{\rm red}\right]
\nonumber\\
&&\times P^i_{\Delta i}
\left[1-i(t_i-t_I)\widehat{H}_{\rm red}\right]
\left|{\bf A}^I_\perp\right>
\nonumber\\
=&&\left<{\bf A}^{I+1}_\perp\right|
\left(1-i\delta t\widehat{H}_{\rm red}\right)
\left|{\bf A}^I_\perp\right>
\\
&&\times\biggl(
    {t_{I+1}-t_i\over\delta t}e_{\Delta_i}\left[A^{I+1}_\perp\right]
    +{t_i-t_I\over\delta t}e_{\Delta_i}\left[A^I_\perp\right]
\biggr).
\nonumber
\end{eqnarray}
In the sum-over-histories formulation, we would choose an ${\bf
A}^\iota_\perp$ most closely corresponding to ${\bf A}_\perp(t_i)$ and
one factor in the indicator function $e_\alpha$ would be
$e_{\Delta_i}[{\bf A}^\iota_\perp]$.  Given the spirit of our lattice
resolution, that is clearly $\overline{\bf A}^I_\perp$.  There is a
discrepancy between the expressions
\begin{mathletters}
\begin{equation}\label{midpt}
e_{\Delta_i}\left[\overline{\bf A}^I_\perp\right]
\end{equation}
and
\begin{equation}\label{ave}
{t_{I+1}-t_i\over\delta t}e_{\Delta_i}\left[{\bf A}^{I+1}_\perp\right]
+{t_i-t_I\over\delta t}e_{\Delta_i}\left[{\bf A}^I_\perp\right],
\end{equation}
\end{mathletters}
even when $t_i=t_I+{\delta t\over 2}$.  However, it can be seen as an
artifact of the lattice; we argued in section \ref{ssec:invar} that as
$\delta t\rightarrow 0$, the difference between ${\bf A}^I_\perp$,
${\bf A}^{I+1}_\perp$, and $\overline{\bf A}^I_\perp$ vanishes like
$(\delta t)^{1/2}$, so we expect both expressions to give the same
results in that limit.  \{It is interesting to note that (\ref{midpt})
and (\ref{ave}) are reminiscent of equations~(20) and (19),
respectively, of Feynman's original paper on path
integrals\cite{feyn}.\}

Here we should consider a moment just what ``projections onto ranges
of ${\bf A_\perp}$'' means, physically.  After all, $\bf A$ is a
gauge-dependent quantity, so it cannot be the expression which
determines the coarse graining in (\ref{latticephase}) independent of
the gauge choice $G$.  The two quantities of interest constructed from
$\bf A$ are $\bf D$ and $\bf B$.  Since $\bf B$, as defined by
(\ref{defB}), depends only on the vector potential $\bf A$ at a single
time, a field configuration $\bf B({\bf x})$ can be determined from a
field configuration $\bf A({\bf x})$ alone.  In fact, the gauge
freedom means there are many configurations of $\bf A$ which lead to
the same $\bf B$ configuration, so there is a one-to-one
correspondence between field configurations $\bf B({\bf x})$ and
gauge-fixed field configurations ${\bf A}_\perp({\bf x})$.

    If we project by values of $\bbox\pi$,
\begin{equation}
P^i_{\Delta_i}
=\int{\cal D}^2\pi^\iota_\perp\left|\bbox\pi^\iota_\perp\right>
e_{\Delta_i}[\bbox\pi^\iota_\perp]\left<\bbox\pi^\iota_\perp\right|,
\end{equation}
we find that to first order in $\delta t$
\widetext
\begin{eqnarray}
\left<{\bf A}^{I+1}_\perp\right|
e^{-i\widehat{H}_{\rm red} (t_{I+1}-t_i)}
&&P^i_{\Delta_i}
e^{-i\widehat{H}_{\rm red} (t_i-t_I)}
\left|{\bf A}^I_\perp\right>
\nonumber\\
&&=
\int{\cal D}^2\pi^\kappa_\perp{\cal D}^2\!A^\iota_\perp
{\cal D}^2\pi^\iota_\perp
\left.\left<{\bf A}^{I+1}_\perp\right|
\bbox\pi^\kappa_\perp\right>
\left<\bbox\pi^\kappa_\perp\left|
{\bf A}^\iota_\perp\right>\right.
\left.\left<{\bf A}^\iota_\perp\right|
\bbox\pi^\iota_\perp\right>
\left<\bbox\pi^\iota_\perp\left|
{\bf A}^I_\perp\right>\right.
\nonumber\\
&&\quad\times
\Biggl\{
    {t_{I+1}-t_i\over\delta t}
    \left(
        1-i\delta t H_{\rm red}\
        \left[
            {{\bf A}^{I+1}_\perp+{\bf A}^\iota_\perp\over 2},
            \bbox\pi^\kappa_\perp
        \right]
    \right)
    e_{\Delta_i}\left[\bbox\pi^\iota_\perp\right]
\nonumber\\
&&\quad\qquad+{t_i-t_I\over\delta t}
    \left(
        1-i\delta t H_{\rm red}\
        \left[
            {{\bf A}^\iota_\perp+{\bf A}^I_\perp\over 2},
            \bbox\pi^\iota_\perp
        \right]
    \right)
    e_{\Delta_i}\left[\bbox\pi^\kappa_\perp\right]
\Biggr\}.
\end{eqnarray}
\narrowtext
If we argue that in the limit $\delta t\rightarrow 0$ we can replace
${\bf A}^\iota_\perp$ with ${\bf A}^I_\perp$ in the first term and
${\bf A}^{I+1}_\perp$ in the second term, we recover the
sum-over-histories expression
\begin{eqnarray}
&&\int{\cal D}^2\pi^\iota_\perp
\left.\left<{\bf A}^{I+1}_\perp\right|
\bbox\pi^\iota_\perp\right>
\left<\bbox\pi^\iota_\perp\left|
{\bf A}^I_\perp\right>\right.
\nonumber\\
&&\times\left(
    1-i\delta t H_{\rm red}\
    \left[
        {{\bf A}^{I+1}_\perp+{\bf A}^I_\perp\over 2},
        \bbox\pi^\iota_\perp
    \right]
\right)
e_{\Delta_i}\left[\bbox\pi^\iota_\perp\right].
\end{eqnarray}

Now we consider physically what a projection onto values of
$\bbox\pi_\perp$ means.  The conjugate momentum $\bbox\pi$ is
gauge-covariant, and in the physical phase space theory directly
accessible, so as long as the regions $\{\Delta_i\}$ are rotationally
invariant in isospace, these are allowed sets of alternatives.
However, dependence of the reduced Hamiltonian (\ref{Hred}) on
$\bbox\pi_\perp$ not just directly but through the fixed form of
$\pi_n$ means that we are actually restricting not $\bbox\pi_\perp$
independently, but $\bbox\pi_\perp$ subject to the constraint ${\bf
D}\cdot\bbox\pi=0$.  This is no cause for alarm, though, since as long
as the coarse graining makes no reference to $\varphi(t_i)$, the
result (\ref{fixK}) [or (\ref{lattfix})] ensures that this is also
the case in the sum-over-histories formulation.

It seems reasonable to assume that, modulo operator ordering
delicacies, a similar correspondence will hold for any combination of
$\bf D$, $\bf B$ and $\bbox\pi$, and {\em the the sum-over-histories
formulation gives the same results as the corresponding reduced phase
space canonical theory} \{referred to in \cite{leshouches} and
elsewhere as ``Arnowitt-Deser-Misner (ADM) quantization''\} {\em for
physical phase space coarse grainings not involving $D_t$.}

\section{Configuration space results} \label{sec:config}

\subsection{Allowed alternatives; overview} \label{ssec:physconf}

    In the sum-over-histories formulation, it is possible to consider
a set of physical alternatives in which the conjugate momenta are not
specified.  The gauge electric field, which was previously described
by $-\bbox\pi$, is now described solely in terms of the potentials as
${\bf E}=-\dot{\bf A}-{\bf D}\varphi$.  While these two definitions of
the gauge electric field are classically equivalent, quantum
mechanical descriptions based on them will in general be inequivalent.
The physical configuration space theory is that in which the gauge
electric field is represented by $\bf E$ and the gauge magnetic field
by $\bf B$.  It has the advantage over the physical phase space
realization that, as described in section \ref{sec:lorentz}, it is
formally manifestly Lorentz-invariant. However, for that very reason,
it will turn out to be not completely consistent with the enforcement
of the Gauss's law constraint.

    In the subset of gauge-invariant alternatives which do not
restrict the momenta (of which the {\em physical} configuration space
alternatives are in turn a subset), the integrals over the momenta in
the path integral (\ref{latticephase}) for the class operator can be
explicitly performed.  We do this in section~\ref{ssec:clatt}, which
gives us a constructive definition of the configuration space path
integral (\ref{formconf}).  For the purposes of the physical
configuration space realization of the NAGT, we could have started
with the formal expression (\ref{formconf}), but approaching it via
the phase space route has enabled us to calculate naturally the
measure for the configuration space path integral.

After constructing the configuration space path integral, we spend the
next two subsections of section~\ref{sec:config} on coarse grainings
by the configuration space constraints.  In section~\ref{SSEC:CONFIGK}
we construct a class of quantities which have the same properties as
those found for the constraints of E\&{}M in \cite{leshouches}: the
only decohering coarse grainings are those which predict that the
constraint is satisfied with 100\% probability.  In
section~\ref{SSEC:EMK}, however, we exhibit a quantity in the abelian
theory of E\&{}M which vanishes in the presence of the Gauss's law
constraint, yet violates this property; there are decohering coarse
grainings which predict nonvanishing values of the quantity with
nonzero probability.

\subsection{Reducing the phase space formulation to the configuration
space formulation}\label{ssec:clatt}

Since the coarse grainings make no reference to the canonical momenta
$\bbox\pi$, we can work in a gauge where $\bbox\pi$ is unrestricted and
perform the Gaussian integrals over the momenta in
(\ref{latticephase}) [using the form of the canonical action density
in (\ref{undo})]:
\widetext
\begin{eqnarray}
\int{\cal D}^3\pi^M\exp
&&\left\{
     -i\delta t\int d^3\!x
     \left[
          {1\over 2}\left(\bbox\pi^M\right)^2
          -\bbox\pi^M\cdot
          \left(
               \dot{\bf A}^M+\overline{\bf D}^M\overline{\varphi}^M
          \right)
     \right]
\right\}\nonumber\\
&&=\prod_{a,\bf x}N_\pi^3\int d^3\pi_a^M({\bf x})\exp
\bigglb(
     -i\delta t\delta^3\!x
     \left\{
          {1\over 2}\left[\bbox\pi_a^M({\bf x})\right]^2
          -\bbox\pi_a^M({\bf x})\cdot
          \left[
               \dot{\bf A}_a^M({\bf x})
               +\left(
                     \overline{\bf D}^M\overline{\varphi}^M
               \right)({\bf x})
          \right]
     \right\}
\biggrb)\nonumber\\
&&=\prod_{a,\bf x}
\left(
     {\delta^3\!x\over 2\pi N_A}\sqrt{2\pi\over i\delta t\delta^3\!x}
\right)^3
\exp
\left\{
     {i\delta t\delta^3\!x\over 2}
     \left[
          \dot{\bf A}_a^M({\bf x})
          +\left(
                \overline{\bf D}^M\overline{\varphi}^M
          \right)_a({\bf x})
     \right]^2
\right\}\nonumber\\
&&=\exp
\left[
     i\delta t\int d^3\!x {1\over 2}\left({\bf E}^M\right)^2
\right]
\prod_{a,\bf x}
\left(
     {\delta^3\!x\over 2\pi i\delta t N_A^2}
\right)^{3/2}
\end{eqnarray}
and rewrite the class operator as
\begin{eqnarray}
\langle\Phi|C_\alpha|\Psi\rangle&=&
{\cal N}\int{\cal D}^4\!A^{J+1}
\Phi^*\!\left[{\bf A}^{J+1},t''\right)\nonumber\\
&&\times
\Bigglb(
     \prod_{M=J}^0\int{\cal D}^4\!A^M\exp
     \left\{i\delta t\int d^3\!x{1\over 2}
          \left[
               \left({\bf E}^M\right)^2-\left({\bf B}^M\right)^2
          \right]
     \right\}
\Biggrb)
\nonumber\\
&&\times\left(\prod_{M=0}^{J+2}
\delta\left[G^M\right]\Delta_{G^M}\right)
\Psi\left[{\bf A}^0,t'\right)e_\alpha[A],\label{latticeconfig}
\end{eqnarray}
\narrowtext
where ${\bf E}^M$ and ${\bf B}^M$ are as given in (\ref{EB}) and we
have defined the normalization constant
\addtocounter{equation}{-1}
\begin{mathletters}
\begin{equation}
{\cal N}=
\left[
     \prod_{a,\bf x}
     \left(
          {\delta^3\!x\over 2\pi i\delta t N_A^2}
     \right)^{3/2}
\right]^{J+1}.
\end{equation}
\end{mathletters}

Looking at (\ref{latticeconfig}), we see that it is a lattice
realization of the formal expression (\ref{formconf}), with the
measure for the configuration space path integral explicitly
calculated.

\subsection{Coarse graining by values of the constraints}
\label{SSEC:CONFIGK}

\subsubsection{Factoring the class operators}

In configuration space, the constraint becomes
\begin{equation}
Q=-{\bf D\cdot E}={\bf D}\cdot\dot{\bf A}+{\bf D}^2\varphi=0
\end{equation}
 and the electric field part of the Lagrangian is
\begin{equation}
\int d^3\!x{1\over 2}{\bf E}^2
=\int d^3\!x{1\over 2}
\left[
     \dot{\bf A}^2+2\dot{\bf A}\cdot{\bf D}\varphi+({\bf D}\varphi)^2
\right].
\end{equation}
 This is most fruitfully simplified by a new gauge, which we dub the
``dotted Coulomb gauge'' (DCG),\footnote{It is possible to show that
we can always reach this gauge, via an argument analogous to that used
in \cite{lee} to show that one can always reach the
Coulomb gauge in a NAGT.} in which
\begin{equation}\label{DCG}
{\bf D}\cdot\dot{\bf A}=0.
\end{equation}
This differs from the Coulomb gauge in which ${\bf D\cdot A}=0$
because the time derivative does not commute with the covariant
gradient $\bf D$. In this gauge, the constraint becomes $Q={\bf
D}^2\varphi=0$ and, after integrating by parts, the electric field
part of the Lagrangian becomes
\begin{equation}
\int d^3\!x{1\over 2}\left[\dot{\bf A}^2+({\bf D}\varphi)^2\right].
\end{equation}
The lattice realization of the gauge condition is
\begin{equation}
G^M=\overline{\bf D}^M\!\cdot\dot{\bf A}^M=0
\end{equation}
for $M=0$ to $J$, which, taken along with
$G^{J+2}=\overline{\varphi}^{J+1}=0$, leaves one hypersurface worth of
gauge conditions $G^{J+1}$ to be specified.  We assume that this is
defined on some hypersurface away from the region examined by our
coarse graining, and ignore it. (It is conventional to assume that it
has been used to ensure that the scalar potential vanishes at spatial
infinity.) The Fadeev-Popov determinant for the DCG can be calculated
from (\ref{fpdet}) to give
\begin{equation}
\Delta_G=\det[-{\bf D}^2\partial_t]=\det[-{\bf D}^2]\det[\partial_t]
\end{equation}
and the class operator is now
\widetext
\begin{eqnarray}
\langle\Phi|C_\alpha|\Psi\rangle&=&
{\cal N}\int{\cal D}^3\!A^{J+1}\det[\partial_t]
\Phi^*\!\left[{\bf A}^{J+1},t''\right)\Psi\left[{\bf A}^0,t'\right)
\nonumber\\
&&\times
\Bigglb(
     \prod_{M=J}^0\int{\cal D}^3\!A^M\exp
     \left\{i\delta t\int d^3\!x{1\over 2}
          \left[
               \left(\dot{\bf A}^M\right)^2-\left({\bf B}^M\right)^2
          \right]
     \right\}
     \delta\left[\overline{\bf D}^M\!\cdot\dot{\bf A}^M\right]
     \det\left[-\left(\overline{\bf D}^M\right)^2\partial_t\right]
\Biggrb)
\nonumber\\
&&\times
\left\{
     \prod_{M=J}^0\int{\cal D}\overline{\varphi}^M\exp
     \left[
          i\delta t\int d^3\!x
          {1\over 2}
          \left(\overline{\bf D}^M\overline{\varphi}^M\right)^2
     \right]
\right\}
\delta\left[G^{J+1}\right]\Delta_{G^{J+1}}
e_\alpha[{\bf A},\varphi].\label{latticeDCG}
\end{eqnarray}
\narrowtext

     The expression (\ref{latticeDCG}) is beginning to factor into two
pieces: a piece depending on the initial and final wave functionals
which involves only the vector potential, and a piece describing the
coarse graining which involves only the scalar potential.  The two
factors which still involve both potentials are the $({\bf
D}\varphi)^2$ term in the exponential and the indicator functional
$e_\alpha[{\bf A},\varphi]$.  We would like to solve the first problem
by changing variables from $\varphi$ to ${\bf D}\varphi$ in the path
integral, but the latter is a vector while the former is only a
scalar.  To construct a scalar corresponding to ${\bf D}\varphi$, we
need to develop some notation.

     First, for the remainder of this section, it will be useful to
consider all unadorned variables to be scalars rather than
four-vectors.  For example, $k=|{\bf k}|=\sqrt{\bf k\cdot k}$.  Now,
we define a nonlocal scalar operator $\nabla=(\bbox\nabla^2)^{1/2}$
via a Fourier transform:
\begin{equation}
\nabla f({\bf x})=\int{d^3\!x'd^3\!k
\over(2\pi)^3} e^{i\bbox{\rm k}\cdot(\bbox{\rm x}-\bbox{\rm x}')}
ikf({\bf x}')
\end{equation}
so that $\nabla^2=\bbox\nabla^2$ is the Laplacian.  (We could have
defined the square root to have the opposite sign, but it would not
substantially change what follows.)  Building on the properties of
this operator, we define an analogous square root for the covariant
Laplacian
\begin{equation}
{\bf D}^2=\bbox\nabla^2
+ig({\bf A}^a\cdot\bbox\nabla+\bbox\nabla\cdot{\bf A}^a)T_a
-g^2({\bf A}^a\cdot{\bf A}^b)T_a T_b
\end{equation}
via an expansion (the convergence of which we do not address) as
follows:
\begin{eqnarray}
D&=&({\bf D}^2)^{1/2}
=\left[\bbox\nabla^2+({\bf D}^2-\bbox\nabla^2)\right]^{1/2}
\nonumber\\
&=&(\bbox\nabla^2)^{1/2}
\left[1+\nabla^{-2}({\bf D}^2-\bbox\nabla^2)\right]^{1/2}
\nonumber\\
&=&\nabla\sum_{n=0}^\infty b_n
\left[\nabla^{-2}({\bf D}^2-\bbox\nabla^2)\right]^n,
\end{eqnarray}
where $\{b_n\}$ are the Taylor expansion coefficients of
$(1+x)^{1/2}$
 and $\nabla^{-2}$ is another non-local operator
\begin{equation}
\nabla^{-2}f({\bf x})=\int{d^3\!x'd^3\!k\over(2\pi)^3}
e^{i\bbox{\rm k}\cdot(\bbox{\rm x}-\bbox{\rm x}')}
\left(-{1\over{\bf k}^2}\right)f({\bf x}')
\end{equation}
defined so that $\bbox\nabla^2\nabla^{-2}=\nabla^{-2}\bbox\nabla^2=1$.

     Now we want to massage the scalar potential part of the action so
that it involves $D\varphi$ rather than ${\bf D}\varphi$.  Integrating
by parts, we see that
\begin{equation}
{1\over 2}\int d^3\!x({\bf D}\varphi)^2=
-{1\over 2}\int d^3\!x\,\varphi\,{\bf D}\cdot{\bf D}\,\varphi=
-{1\over 2}\int d^3\!x\,\varphi D^2\varphi
\end{equation}
and now we need to move one of the $D$ operators back to the left.  It
is straightforward to show (by expanding $\alpha$ and $\beta$ in
Fourier transforms) that
\begin{equation}
\int d^3\!x\,\alpha({\bf x})\nabla\beta({\bf x})
=\int d^3\!x(\nabla\alpha)({\bf x})\beta({\bf x})
\end{equation}
(which is the opposite sign from the integration by parts involving
$\bbox\nabla$) and
\begin{equation}
\int d^3\!x\,\alpha({\bf x})\nabla^{-2}\beta({\bf x})
=\int d^3\!x(\nabla^{-2}\alpha)({\bf x})\beta({\bf x}).
\end{equation}
Using those two results, along with
\begin{eqnarray}
\int d^3\!x\,\alpha({\bf x})&&({\bf D}^2-\bbox\nabla^2)\beta({\bf x})
\nonumber\\
&&=\int d^3\!x
\left[({\bf D}^2-\bbox\nabla^2)\alpha({\bf x})\right]\beta({\bf x})
\end{eqnarray}
(which follows from the integration by parts procedures for
$\bbox\nabla$ and $\bf D$) one can show that
\begin{equation}
-{1\over 2}\int d^3\!x\,\varphi D^2\varphi=
-{1\over 2}\int d^3\!x(D\varphi)^2.
\end{equation}
 If we define\footnote{The factor of $i$ is necessary to make ${\cal
E}(\bf x)$ a real quantity.  In E\&{}M, this change of variables is
just changing to ${\cal E}=-iE_L$ where $E_L$ is the (scalar)
longitudinal part of the electric field.}
\begin{equation}
{\cal E}^M=i\overline{D}^M\overline{\varphi}^M,
\end{equation}
the class operator becomes
\widetext
\begin{eqnarray}
\langle\Phi|C_\alpha|\Psi\rangle&=&
{\cal N}\int{\cal D}^3\!A^{J+1}\det[\partial_t]
\Phi^*\!\left[{\bf A}^{J+1},t''\right)\Psi\left[{\bf A}^0,t'\right)
\nonumber\\
&&\times
\Bigglb(
     \prod_{M=J}^0\int{\cal D}^3\!A^M\exp
     \left\{i\delta t\int d^3\!x{1\over 2}
          \left[
               \left(\dot{\bf A}^M\right)^2-\left({\bf B}^M\right)^2
          \right]
     \right\}
     \delta\left[\overline{\bf D}^M\!\cdot\dot{\bf A}^M\right]
     \det\left[i\overline{D}^M\partial_t\right]
\Biggrb)\nonumber\\
&&\times
\left\{
     \prod_{M=J}^0\int{\cal D}{\cal E}^M\exp
     \left[
          i\delta t\int d^3\!x{1\over 2}\left({\cal E}^M\right)^2
     \right]
\right\}
\delta\left[G^{J+1}\right]\Delta_{G^{J+1}}
e_\alpha[{\bf A},{\cal E}].
\end{eqnarray}

     Now the only obstacle to factorization of the class operator is
the indicator functional $e_\alpha$.  If we coarse grain by some
temporal and spatial average of the constraint
\begin{equation}
Q={\bf D}^2\varphi=-iD{\cal E},
\end{equation}
the indicator functional for a class in which this average lies in the
range $\Delta$ (which, since $Q$ is an isovector, is a region in
isospace which is mapped onto itself by gauge transformations) is
(letting $n=\sum_a 1$ be the dimension of the adjoint representation
of the gauge group, and keeping in mind that $f$ is a complex
isovector quantity)
\begin{equation}
e_\Delta
=\int\limits_\Delta d^{2n}\!f\,\delta(f-\langle -iD{\cal E}\rangle),
\end{equation}
 which depends on $\bf A$ via the operator $D$.  If, however, we
coarse grain by values of $iD^{-1}Q$, which classically should vanish
whenever $Q$ does, we are coarse graining by $\cal E$, $e_\Delta$ is
independent of $\bf A$, and we can perform the following manipulation:
\begin{eqnarray}
\langle\Phi|C_\Delta|\Psi\rangle&=&
{\cal N}{\cal C}_\Delta\int{\cal D}^3\!A^{J+1}\det[\partial_t]
\Phi^*\!\left[{\bf A}^{J+1},t''\right)\Psi\left[{\bf A}^0,t'\right)
\nonumber\\
&&\times
\Bigglb(
     \prod_{M=J}^0\int{\cal D}^3\!A^M\exp
     \left\{i\delta t\int d^3\!x{1\over 2}
          \left[
               \left(\dot{\bf A}^M\right)^2-\left({\bf B}^M\right)^2
          \right]
     \right\}
     \delta\left[\overline{\bf D}^M\!\cdot\dot{\bf A}^M\right]
     \det\left[i\overline{D}^M\partial_t\right]
\Biggrb)\nonumber\\
&&\times\delta\left[G^{J+1}\right]\Delta_{G^{J+1}},
\end{eqnarray}
\narrowtext
where
\begin{equation}\label{cD}
{\cal C}_\Delta=
\left\{
     \prod_{M=J}^0\int{\cal D}{\cal E}^M\exp
     \left[
          i\delta t\int d^3\!x{1\over 2}\left({\cal E}^M\right)^2
     \right]
\right\}
e_\Delta[{\cal E}].
\end{equation}
This means that
\begin{equation}
D(\Delta,\Delta')={{\cal C}^*_\Delta{\cal C}_{\Delta'}
\over|{\cal C}_u|^2}
\end{equation}
and we can apply an argument from section VI.4 of \cite{leshouches}:
when the decoherence functional factors in this way, the only way the
off-diagonal elements can vanish is if only one of the $\{{\cal
C}_\Delta\}$ is non-zero.  In that case, one diagonal element of the
decoherence functional is unity and all the others vanish, which
corresponds to a definite prediction of that alternative (100\%
probability).  Thus we have the result: {\em coarse grainings of
$iD^{-1}(-{\bf D\cdot E})$ in configuration space fall into two
categories: either they yield a definite prediction of a single
alternative, or they fail to decohere.} In the former case, we expect
that the predicted alternative will be the one consistent with
constraint $Q=0$, but this argument itself does not settle the issue.
However, the conjecture seems very likely given that the integrand in
the expression (\ref{cD}) for ${\cal C}_\Delta$ is stationary about
${\cal E}=0$, which would seem to make the alternative including
${\cal E}=0$ the one most likely to have a non-vanishing ${\cal
C}_\Delta$.

\subsubsection{A decohering example}\label{SSSEC:CONFIGK}

    We now present explicit calculation of ${\cal C}_\Delta$ for one
choice of the average $\langle{\cal E}\rangle$ which verifies both
that coarse grainings of the first class exist and that the
alternative predicted is (in this case) indeed the one consistent with
the constraint.  The demonstration is analogous to that used in
section VI.4 of \cite{leshouches} for E\&{}M, and the specialization
of the present result (\ref{eqA}) to the abelian case is in fact a
more accurate version of equation (VI.4.12) therein.

We coarse grain by an average $\langle{\cal E}\rangle$ over modes so
that the indicator functional is
\begin{equation}\label{eD}
e_\Delta[{\cal E}]
=\int\limits_\Delta d^{2n}\!f\delta(f-\langle{\cal E}\rangle).
\end{equation}
The average
$\langle{\cal E}\rangle$ is defined to be over a time interval $\Delta
t$ and a group of modes in spatial frequency space $\Delta\!^3\!k$.  We
refer to this group of modes as $\Omega$, which we also use for the
mode volume ($\Omega=\Delta t\Delta\!^3\!k$), so that the average is
\begin{equation}\label{modeave}
\langle{\cal E}\rangle={1\over\Omega}\int\limits_\Omega dt\, d^3\!k\,
{\cal E}_{\bf k}(t)
={1\over\Omega}\sum_{M\in\Omega}\delta t\int\limits_\Omega d^3\!k\,
{\cal E}^M_{\bf k},
\end{equation}
where ${\cal E}_{\bf k}$ is the Fourier transform
\begin{equation}\label{fourE}
{\cal E}^M_{\bf k}=\int{d^3\!x\over(2\pi)^{3/2}}
e^{-i\bbox{\rm k}\cdot\bbox{\rm x}}{\cal E}^M({\bf x}).
\end{equation}

    For this coarse graining, the calculation in
appendix~\ref{app:NAK} gives
\begin{equation}\eqnum{\protect\ref{eqA}}
{\cal C}_\Delta={\cal K}'\int\limits_\Delta d^{2n}\!f e^{i\Omega|f|^2},
\end{equation}
where ${\cal K}'$ is a constant. The integrand is an ``imaginary
Gaussian'' of width $1/\sqrt{2\Omega}$; For $|f|^2\gtrsim 1/2\Omega$,
the integrand will oscillate rapidly and the contributions to the
integral will cancel out.  This means that if we average over a large
enough group of modes $\Omega$ that the region
\begin{equation}
\left|{1\over\Omega}\int\limits_\Omega dt\, d^3\!k\,
{\cal E}_{\bf k}(t)\right|^2
\lesssim{1\over 2\Omega}
\end{equation}
is contained in a single bin $\Delta$, that will correspond to the
only non-negligible ${\cal C}_\Delta$, and we will have a definite
prediction that the configuration space constraint is satisfied to
that accuracy.

     This result is less comforting than the abelian one, since our
alternatives were defined not by the usual configuration space
constraint $Q=-\bf D\cdot E$ but a nonlocal function of it.  In
E\&{}M, no one would object to analogously coarse graining by the
longitudinal component of the electric field rather than its
divergence, but in that case the relationship between them does not
involve the other components of $\bf A$ so a similar factorization can
be performed on $-\bbox\nabla\cdot\bf E$ as on $E_L$.  However, in a
NAGT, coarse graining by $Q=-iD{\cal E}$ tangles up ${\cal E}$ and
$\bf A$.  Even in E\&{}M, we run into this problem if we coarse grain
by quantities which involve both ${E_L}$ and ${\bf A}_T$.  We examine
one such coarse graining in the next section.

\subsection{Coarse graining E\&{}M by quantities proportional to the
constraint} \label{SSEC:EMK}

    We showed in the previous section that coarse grainings by values
of $\cal E$ in a NAGT (or $E_L$ in E\&{}M) could only decohere in
cases where they led to a definite prediction.  The demonstration does
not work for coarse grainings by $-iD{\cal E}$ (or $f[E_L,{\bf B}]$ in
E\&{}M).  We will now exhibit such a coarse graining in E\&{}M which
decoheres, but predicts non-zero probabilities for more than one
alternative, thus verifying that the property described in the
previous section does not always hold.

    The physical process believed to be responsible for decoherence in
most practical situations of everyday life results when (see
\cite{classeq} for details and a bibliography of prior work) the
``system'' of interest is coupled
to an ``environment''.  The ``environment''
is not measured, but carries away phase information which causes sets
of alternatives describing the ``system'' to decohere.  The present
situation here is similar, but with the following differences.  The
``system'' variables ${\bf A}$ are coupled to the ``environment''
variables ${\cal E}$ not by the action, but by the coarse graining
itself, and here it is the initial state rather than the coarse
graining which is independent of the ``environment'' ${\cal E}$.  But
as we shall see, this is still a mechanism which can produce
decoherence of a sort different than that seen in the previous
section, and lead to more than one alternative having non-zero
probability.  For our purposes, it will be most useful to consider
coarse grainings by functionals of $E_L$ and ${\bf B}$ in the abelian
gauge theory of electromagnetism.  (Although we will briefly mention,
at the end of the section, a similar result in another theory to
illustrate the generality of the mechanism described here.)

    In the abelian theory, the dotted Coulomb gauge (\ref{DCG}) is
equivalent to the Coulomb gauge $A_L=0$ and (\ref{latticeDCG}) becomes
\widetext
\begin{eqnarray}
\langle\Phi|C_\alpha|\Psi\rangle&=&
{\cal N}\int{\cal D}^2\!A_T^{J+1}\det[\partial_t]
\Phi^*\!\left[{\bf A}_T^{J+1},t''\right)\Psi\left[{\bf A}_T^0,t'\right)
\nonumber\\
&&\times
\Bigglb(
     \prod_{M=J}^0\int{\cal D}^2\!A_T^M\exp
     \left\{i\delta t\int d^3\!x{1\over 2}
          \left[
               \left(\dot{\bf A}_T^M\right)^2
               -\left(\bbox\nabla\times\dot{\bf A}_T^M\right)^2
          \right]
     \right\}
     \det[-\nabla]
\Biggrb)\nonumber\\
&&\times
\left\{
     \prod_{M=J}^0\int{\cal D}\overline{\varphi}^M\exp
     \left[
          i\delta t\int d^3\!x{1\over 2}
          \left(\bbox\nabla\overline{\varphi}^M\right)^2
     \right]
\right\}
e_\alpha[{\bf A}_T,\varphi].\label{lattEM}
\end{eqnarray}
    For the purposes of E\&{}M, we need not concern ourselves with the
details of the lattice approximation, because the reduced Hamiltonian
is free from the operator ordering ambiguities discussed in
section~\ref{ssec:lattred}.\footnote{For example, in the axial gauge
(\ref{ax}), the functional defined in (\ref{pin}) becomes $\pi_n[{\bf
A}_\perp,\bbox\pi_\perp]
=-\partial_n^{-1}\bbox\nabla_\perp\cdot\bbox\pi_\perp$, which is
independent of ${\bf A}_\perp$ so that ${1\over 2}(\pi_n[{\bf
A}_\perp,\bbox\pi_\perp])^2$ can be unambiguously converted into an
operator expression.} This means that any choice of operator ordering
convention gives the same reduced Hamiltonian, and in light of the
discussion in footnote~\ref{fn:opord}, page~\pageref{fn:opord} this
means that different lattice realizations of the path integral will be
equivalent.  We are thus justified in working with the formal
equivalent of (\ref{lattEM}):
\begin{equation}
\langle\Phi|C_\alpha|\Psi\rangle
=\Delta_G\int\limits_\alpha{\cal D}^2\!A_T\,{\cal D}\varphi\,
\Phi^*\!\left[{\bf A}_T^{J+1},t''\right)
\exp
\left\{i\int d^4\!x{1\over 2}
     \left[
          \left(\dot{\bf A}_T\right)^2
          -\left(\bbox\nabla\times\dot{\bf A}_T\right)^2
          +\left(\bbox\nabla\varphi\right)^2
     \right]
\right\}
\Psi\left[{\bf A}_T^0,t'\right).
\end{equation}
\narrowtext

 Since we can observe that the ``physical degrees of freedom'' upon
which the wave functionals depend are just the transverse components of
${\bf A}$, it is useful to factor out the wave functionals and write
\begin{eqnarray}
\langle\Phi|C_\alpha|\Psi\rangle=
\int&&{\cal D}^2\!A''_T\,{\cal D}^2\!A'_T\,
\Phi^*\!\left[{\bf A}_T'',t''\right)
\nonumber\\
&&\times
C_\alpha[{\bf A}''_T t''|{\bf A}'_T t')
\Psi\left[{\bf A}'_T,t'\right),\label{fwf}
\end{eqnarray}
where\footnote{Here we abuse the mismatched-parentheses notation
slightly. $C_\alpha[{\bf A}''_T t''|{\bf A}'_T t')$ is a functional of
${\bf A}''$ and ${\bf A}'$ and a function of $t''$ and $t'$.}
\begin{equation}
C_\alpha[{\bf A}''_T t''|{\bf A}'_T t')
=\Delta_G\int\limits_{{\bf A}'_T\alpha{\bf A}''_T}
{\cal D}^2\!A_T\,{\cal D}\varphi\,e^{i\int_{t'}^{t''}\!dt L}.
\end{equation}
If we write the initial and final conditions as
\begin{mathletters}
\begin{eqnarray}
\rho''_{t''}[{\bf A}''_{T2},{\bf A}''_{T1}] &=&\sum_i
\Phi_i^*\!\left[{\bf A}_{T2}'',t''\right) p''_i
\Phi_i\left[{\bf A}_{T1}'',t''\right)
\\
\rho'_{t'}[{\bf A}'_{T2},{\bf A}'_{T1}] &=&\sum_j
\Psi_j^*\!\left[{\bf A}_{T2}',t'\right)p'_j
\Psi_j\left[{\bf A}_{T1}',t'\right),
\end{eqnarray}
\end{mathletters}
we have, from (\ref{fwf}) and (\ref{dec}),
\begin{eqnarray}
D(\alpha,\alpha')&\propto&
\int{\cal D}^2\!A_{T2}''{\cal D}^2\!A_{T1}''
{\cal D}^2\!A_{T2}'{\cal D}^2\!A_{T1}'
\nonumber\\
&&\times\rho''_{t''}[{\bf A}''_{T2},{\bf A}''_{T1}]
C_\alpha[{\bf A}''_{T1} t''|{\bf A}'_{T1} t')
\nonumber\\
&&\times\rho'_{t'}[{\bf A}'_{T1},{\bf A}'_{T2}]
C^*_{\alpha'}[{\bf A}''_{T2} t''|{\bf A}'_{T2} t'),
\label{Dsep}
\end{eqnarray}
where we have established the useful convention that $\propto$
indicates a proportionality constant which is the same for all classes
and thus
can be absorbed into the normalization.  The quantity we choose to
define our alternatives is
\begin{equation}
g[{\bf B}^\iota]\bigl|\langle E_L\rangle\bigr|^2,
\end{equation}
where $g[{\bf B}^\iota]$ is a functional (which we take to be
positive semidefinite for reasons to become clear later) of the
magnetic field configuration ${\bf B^\iota}$ on some time slice $t_i$,
and $\langle\cdot\rangle$ indicates an average over some mode volume
(i.e., an average over wavenumber and time\footnote{Since it
involves a time average, this sort of alternative is not accessible in
a standard operator-and-state formulation of quantum mechanics.}).
The indicator function for this quantity to lie in some interval
$\Delta$ is
\begin{eqnarray}
e_\Delta&=&\int\limits_\Delta df
\delta\!\left(f-g[{\bf B}^\iota ]
\bigl|\langle E_L\rangle\bigr|^2\right)
\nonumber\\
&=&\int\limits_\Delta df\int db\, da\,
\delta(f-ba)\delta(b-g[{\bf B}^\iota ])
\nonumber\\
&&\qquad\qquad\times
\delta\!\left(a-\bigl|\langle E_L\rangle\bigr|^2\right),
\end{eqnarray}
which allows us to write
\begin{eqnarray}
C_\Delta[{\bf A}''_T t''|&&{\bf A}'_T t')
\nonumber\\
&&=\int\limits_\Delta df\int db\, da\delta(f-ba){\cal A}(a)
{\cal B}\left[{\bf A}''_T,{\bf A}'_T,b\right)
\nonumber\\
&&=\int\limits_\Delta df\int {db\over |b|}{\cal A}
\left({f\over b}\right)
{\cal B}\left[{\bf A}''_T,{\bf A}'_T,b\right),
\label{Csep}
\end{eqnarray}
where
\addtocounter{equation}{-1}
\begin{mathletters}
\begin{eqnarray}
{\cal A}(a)=\int&&{\cal D}\varphi\exp
\left[
	i\int_{t'}^{t''}\!dt\int d^3\!x{1\over 2}
	\left(\bbox\nabla\varphi\right)^2
\right]
\nonumber\\
&&\times\delta\left(a-|\langle\nabla\varphi\rangle|^2\right)
\det[-\nabla]
\end{eqnarray}
and
\begin{eqnarray}
&&{\cal B}\left[{\bf A}''_T,{\bf A}'_T,b\right)
\nonumber\\
&&=\int\limits_{{\bf A}'_T{\bf A}''_T}{\cal D}^2\!A_T
\nonumber\\
&&\qquad
\times\exp
\left\{
    i\int_{t'}^{t''}\!dt\int d^3\!x{1\over 2}
    \left[
        \left(
             \dot{\bf A}_T
        \right)^2
        -
        \left(
            \bbox\nabla\times{\bf A}_T
        \right)^2
    \right]
\right\}
\nonumber\\
&&\qquad
\times\delta(b-g[\bbox\nabla\times{\bf A}^\iota _T]).
\end{eqnarray}
\end{mathletters}
Writing the average over a group of modes as
\begin{equation}
\langle\nabla\varphi\rangle=
{1\over\Omega}\int\limits_\Omega dt\, d^3\!k
[ik\overline{\varphi}_{\bf k}(t)],
\end{equation}
a calculation analogous to the one in appendix~\ref{app:NAK} tells us
that
\begin{eqnarray}
{\cal A}(a)&\propto&\int{\cal D}\Upsilon\exp
\left(
    i\Omega\sum_{\sigma\in\Omega}|\Upsilon_\sigma|^2
\right)
\delta\left(a-|\Upsilon_0|^2\right)
\nonumber\\
&\propto&
\int d\Upsilon^{\rm R}_0 d\Upsilon^{\rm I}_0\exp
\left(
    i\Omega|\Upsilon_0|^2
\right)
\delta\left(a-|\Upsilon_0|^2\right)
\nonumber\\
&\propto&
e^{i\Omega a}\Theta(a),\label{Ares}
\end{eqnarray}
where $\Theta(a)$ is the Heaviside step function. Meanwhile, we can
write ${\cal B}$ as\footnote{The attentive reader may notice that we
are implicitly expressing our coarse graining in terms of ${\bf
A}^\iota$ as though it corresponded to $\bf A$ on a single lattice
slice $\left({\bf A}^I\right)$ rather than an average
$\left(\overline{\bf A}^I\right)$, as we were instructed to do in
section \ref{ssec:lattice}.  This is not a problem, because, as
discussed earlier, the operator ordering ambiguities that led us to
make the distinction between the two are not present in E\&{}M.}
\widetext
\begin{equation}\label{calB}
{\cal B}\left[{\bf A}''_T,{\bf A}'_T,b\right)
=\int{\cal D}^2\!A^\iota _T\,
{\cal G}\left[{\bf A}''_T t''|{\bf A}^\iota _T t_i\right)
\delta(b-g[\bbox\nabla\times{\bf A}^\iota _T])\,
{\cal G}\left[{\bf A}^\iota _T t_i|{\bf A}'_T t'\right),
\end{equation}
where
\begin{equation}
{\cal G}\left[{\bf A}''_T t''|{\bf A}'_T t'\right)
=\int\limits_{{\bf A}'_T{\bf A}''_T}{\cal D}^2\!A_T
\exp
\left\{
    i\int_{t'}^{t''}dt\int d^3\!x{1\over 2}
    \left[
        \left(
             \dot{\bf A}_T
        \right)^2
        -
        \left(
            \bbox\nabla\times{\bf A}_T
        \right)^2
    \right]
\right\}
\end{equation}
is the propagator for the ${\bf A}_T$ sector of the theory.  By
Fourier transforming the spatial dependence of ${\bf A}_T$, $\cal G$
can be seen to be the propagator for a harmonic oscillator whose
natural frequency depends on the wave number ${\bf k}$ of the mode.
Equation (\ref{calB}) allows us to write the dependence implied by
(\ref{Dsep}) and (\ref{Csep}) of $D(\Delta_2,\Delta_1)$ on the initial
and final conditions as
\begin{eqnarray}
&&\int{\cal D}^2\!A_{T2}''{\cal D}^2\!A_{T1}''
{\cal D}^2\!A_{T2}'{\cal D}^2\!A_{T1}'
\rho''_{t''}[{\bf A}''_{T2},{\bf A}''_{T1}]
{\cal B}\left[{\bf A}''_{T1},{\bf A}'_{T1},b_1\right)
\rho'_{t'}[{\bf A}'_{T1},{\bf A}'_{T2}]
{\cal B}^*\!\left[{\bf A}''_{T2},{\bf A}'_{T2},b_2\right)
\nonumber\\
&&=
\int{\cal D}^2\!A^\iota _{T2}{\cal D}^2\!A^\iota _{T1}
\rho''_{t_i}[{\bf A}^\iota _{T2},{\bf A}^\iota _{T1}]
\delta(b_1-g[\bbox\nabla\times{\bf A}^\iota _{T1}])
\rho'_{t_i}[{\bf A}^\iota _{T1},{\bf A}^\iota _{T2}]
\delta(b_2-g[\bbox\nabla\times{\bf A}^\iota _{T2}]),
\label{fnord}
\end{eqnarray}
\narrowtext where we have used the propagator $\cal G$ to transform
$\rho'_{t'}$ to $\rho'_{t_i}$ and $\rho''_{t''}$ to $\rho''_{t_i}$.
If the final state is one of future indifference:
\begin{equation}
\rho''[{\bf A}''_{T2},{\bf A}''_{T1}]
\propto\delta[{\bf A}''_{T2}-{\bf A}''_{T1}]
\end{equation}
(which is preserved by the propagator), (\ref{fnord}) becomes
proportional to
\begin{eqnarray}
\int&&{\cal D}^2\!A^\iota_T\,
\delta(b_1-g[\bbox\nabla\times{\bf A}^\iota _T])
\nonumber\\
&&\qquad\times
\rho'_{t_i}[{\bf A}^\iota_T,{\bf A}^\iota _T]
\delta(b_2-g[\bbox\nabla\times{\bf A}^\iota _T])
\nonumber\\
&&=\delta(b_2-b_1)\,p(b_1),
\end{eqnarray}
where
\begin{equation}\label{probb}
p(b)=\int{\cal D}^2\!A^\iota _T\,
\delta(b-g[\bbox\nabla\times{\bf A}^\iota _T])
\rho'_{t_i}[{\bf A}^\iota _T,{\bf A}^\iota _T].
\end{equation}

Combining (\ref{Dsep}) and (\ref{Csep}) with the expression for $\cal
A$ in (\ref{Ares}) and this result concerning ${\cal B}$, we have
\begin{eqnarray}
&&D(\Delta,\Delta')
\\
&&\qquad\propto
\int\limits_\Delta df\int\limits_\Delta df'
\int{db\over b^2}p(b)e^{i\Omega(f-f')/b}\,
\Theta\left(f\over b\right)\Theta\left(f'\over b\right).
\nonumber
\end{eqnarray}
With the condition that $g[{\bf B}^\iota]$ is everywhere
non-negative, we see from (\ref{probb}) that $p(b)$ vanishes for
negative $b$ and the step functions above become
$\Theta(f)\Theta(f')$.  If we define the regions \{$\Delta$\} to cover
the positive real axis, we can drop the step functions to give
\begin{equation}\label{DG}\label{DGmath}
D(\Delta,\Delta')\propto
\int\limits_\Delta df\int\limits_\Delta df'\, G(f-f'),
\end{equation}
where
\addtocounter{equation}{-1}
\begin{mathletters}
\begin{equation}\label{Gy}
G(y)=\int_0^\infty{db\over b^2}p(b)e^{i\Omega y/b}.
\end{equation}
\end{mathletters}
 Note that since $G(f-f')$ depends only on the difference between $f$
and $f'$, no value of $f$ is preferred over any other.  In particular,
if the bins $\{\Delta\}$ are all the same size, $D(\Delta,\Delta')$
depends only on the relative separation of $\Delta$ and $\Delta'$, not
their absolute location.  This means that if there is decoherence,
(\ref{DG}) predicts that the measured quantity is equally likely to
have any value.

    It is possible to choose the $p(b)$ (which is determined by the
initial conditions) to produce at least weak decoherence.  For
example, let $p(b)$ be a Gaussian in $1/b$:
\begin{equation}\label{pbgauss}
p(b)=A_\lambda\Theta(b)\Theta(\lambda^{-1}-b)e^{-1/2\sigma^2b^2},
\end{equation}
where $A_\lambda$ is a cutoff-dependent normalization given by
\addtocounter{equation}{-1}
\begin{mathletters}
\begin{equation}
A_\lambda^{-1}=\int_0^{\lambda^{-1}}db\,e^{-1/2\sigma^2b^2}
<{\sigma\sqrt{2\pi}\over\lambda^2}
\end{equation}
\end{mathletters}
to ensure $\int_0^\infty p(b)=1$.

    If $\lambda$ is small enough, the leading terms in the decoherence
functional will not depend on it.  If the real parts of the
off-diagonal elements of the decoherence functional are much less than
the diagonal elements, the coarse graining will exhibit approximate
weak decoherence [cf.\ (\ref{weak})] The calculation in
appendix~\ref{app:pb} shows for bins of equal size $\Delta$ that, to
lowest order in $e^{-(\Omega\sigma\Delta)^2/2}$,
\begin{equation}
{\mathop{\rm Re}\nolimits D(J+\Delta{J},J)\over D(J,J)}
\lesssim{\exp[-(\Omega\sigma\Delta)^2(|\Delta{J}|-1)^2/2]
\over\Omega\sigma\Delta\sqrt{2\pi}}.
\end{equation}

So $D(J,J\pm 1)$ is suppressed by a factor of
$(\Omega\sigma\Delta)^{-1}$ relative to $D(J,J)$, while all the other
elements of the decoherence functional are exponentially suppressed.
In general, we expect this sort of result if $\mathop{\rm Re}\nolimits
G(y)$ falls off on a scale which is small compared to $\Delta$ [which
should in general be determinable from a steepest descents evaluation
of (\ref{Gy}).]  Schematically (Fig.~\ref{fig:dropoff}), if
$\mathop{\rm Re}\nolimits G(y)$ becomes negligible for
$|y|\gtrsim\delta$, with $\delta\le\Delta$, the integral for
$\mathop{\rm Re}\nolimits D(J,J+\Delta{J})$ for $|\Delta{J}|\ge 2$
will include none of the region for which $G(y)$ is significant.  The
area of the region in the integral for $\mathop{\rm Re}\nolimits
D(J,J\pm1)$ for which $G(y)$ is significant is $\delta^2/2$, while
that for $D(J,J)$ is $2\Delta\delta-\delta^2$, so $\mathop{\rm
Re}\nolimits D(J,J\pm1)$ is suppressed by a factor of $\delta/\Delta$.

  This means that if $\Omega\sigma\Delta\gg 1$, {\em this coarse
graining by $g[{\bf B}^\iota ]\bigl|\langle E_L\rangle\bigr|^2$
decoheres weakly for the initial condition (\ref{pbgauss}) and the
final condition of future indifference, and there is an equal
probability for the value to fall into any of the evenly spaced bins.}
A curious corollary is that if we coarse grain by combining bins $0$
through $J_0-1$ into one alternative $c_<$, corresponding to $g[{\bf
B}^\iota ]|\langle E_L\rangle|^2<J_0\Delta$, and all the bins $J_0$
and up into another, corresponding to $g[{\bf B}^\iota ]\bigl|\langle
E_L\rangle\bigr|^2>J_0\Delta$, we find that since $p_<$ is a sum of
$J_0$ equal terms and $p_>$ is an infinite sum of the same terms,
$p_<=0$ and $p_>=1$ for any finite $J_0$, a definite prediction that
$g[{\bf B}^\iota]|\langle E_L\rangle|^2>J_0\Delta$.  This sort of
phenomenon is common in the use of path integral methods (for another
example, see \cite{har88}) and is related to the non-differentiability
of Brownian paths.

    Finally, let us comment on the significance of this result.  If we
coarse grained by values of the corresponding phase space quantity,
$g[{\bf B}^\iota]|\langle \pi_L\rangle|^2$, (\ref{fixK}) would ensure
that we found a definite prediction that it vanished.  Thus the phase
space and configuration space theories make different predictions.
There are well-documented examples in generalized quantum mechanics
where analogous phase space and configuration space coarse grainings
lead to different decoherence functionals (for example, in Sections
V.4.2 and VI.4 of\cite{leshouches}), but in those cases, there was a
coarse graining by momentum which decohered while coarse graining by
the equivalent quantity in terms of velocity did not. When the coarse
graining by velocity decohered, it agreed with the coarse graining by
momentum. The present result is the first case known to the author of
corresponding phase space and configuration space coarse grainings,
both of which decohere, but which give conflicting probabilities.

    This result is not limited to constrained theories.  Another
system in which similar phenomena can occur is the non-relativistic
quantum mechanics of a free particle with two degrees of freedom and
an independent harmonic oscillator.  In that case, one coarse grains
by the product of some function of the position of the harmonic
oscillator at one instant of time with the square of a time average of
the velocity of the free particle.  If the initial state is a
zero-momentum eigenstate of the free particle tensored with a suitable
state of the harmonic oscillator, one finds equal probability of any
alternative, even though the corresponding phase space coarse
graining yields a definite prediction that the quantity vanishes.

\section{A few words about Lorentz invariance}\label{sec:lorentz}

    Since our implementation of the sum over histories for the
generalized quantum mechanics of a NAGT has relied rather heavily on a
division into time and space, it is worth mentioning how little the
formal theory really does to single out a preferred Lorentz frame.
The phase space theory is of course not Lorentz-invariant, as the
conjugate momenta are defined with respect to a particular time.
Since $-\bbox\pi$ and ${\bf B}$ are treated differently, it is not
possible to combine them into a field strength tensor which Lorentz
transforms appropriately. This is the source of the apparent asymmetry
between different components of the equations of motion
\begin{equation}
D_\mu G^{\mu\nu}=0;
\end{equation}
the constraints hold identically, while the others do not.

    However, the formal configuration space theory (and not just the
``physical'' configuration space coarse grainings defined in section
\ref{ssec:physconf}) {\em can} be cast into a form which is manifestly
Lorentz-invariant.  In the formal configuration space expression
\begin{eqnarray}\label{formconf2}
\langle\Phi|C_\alpha|\Psi\rangle
&=&\int\limits_\alpha{\cal D}^4\!A\,\Phi^*\![{\bf A}'',t'')
\delta[G]\Delta_G[A]
\\
&&\times
\exp
\left(
    -i\int_{t'}^{t''}\!dt\int\!d^3\!x{1\over 4}G^a_{\mu\nu}G_a^{\mu\nu}
\right)
\Psi[{\bf A}',t'),
\nonumber
\end{eqnarray}
$\bf E$ and $\bf B$ are treated on equal footing from a spacetime
point of view as part of the tensor $G^{\mu\nu}$.  Lorentz invariance
is broken in two ways, both concerning the initial and final wave
functionals $\Psi$ and $\Phi$.  First, they are attached on surfaces
of constant co\"{o}rdinate time rather than arbitrary spacelike
surfaces; second, the operator constraints (\ref{Psicons}) on $\Psi$
and $\Phi$ treat $\varphi=A^0$ and ${\bf A}=A^i{\bf e}_i$ unequally.
In this section, we demonstrate that these two problems are related to
one another, and show how the conditions satisfied by the wave
functionals can be related to the surfaces on which they are
evaluated.

    We can generalize (\ref{formconf2}) to arbitrary (spacelike)
initial and final surfaces in the straightforward manner:
\begin{eqnarray}
\langle\Phi|C_\alpha|\Psi\rangle
=\int\limits_\alpha&&{\cal D}^4\!A\,
\Phi^*\!\left[A^{(\sigma'')},\sigma''\right)
\delta[G]\Delta_G[A]
\nonumber\\
&&\times
e^{i\int_{\sigma'}^{\sigma''}\!d^4\!x{\cal L}(x)}
\Psi[A^{(\sigma')},\sigma'),
\label{formform}
\end{eqnarray}
where $A^{(\sigma)}$ is the restriction of the function $A(x)$ (here
the four-vector potential, but the definition will apply to any
function defined over spacetime) onto the three-surface $\sigma$, and
the integral for the action is over the region bounded by $\sigma'$
and $\sigma''$.  Using the sum over all histories to define a
propagator\footnote{Again, the mixed-parentheses notation is stretched
somewhat here; ${\cal G}\left[A^{(\sigma'')}\sigma''\right|
\left.A^{(\sigma')}\sigma'\right)$ is a functional of the potentials
$A^{(\sigma')}$ and $A^{(\sigma'')}$ and a ``function'' of the
spacelike surfaces $\sigma'$ and $\sigma''$.}
\begin{eqnarray}
&&{\cal G}\left[A^{(\sigma'')}\sigma''\right|
\left.A^{(\sigma')}\sigma'\right)
\nonumber\\
&&=\!\!\!\int\limits_{A^{(\sigma')}A^{(\sigma'')}}\!\!\!{\cal D}^4\!A\,
\delta[G]\Delta_G[A]e^{i\int_{\sigma'}^{\sigma''}\!d^4\!x{\cal L}(x)},
\label{formprop}
\end{eqnarray}
we can go from a wave functional $\Psi$ defined on one spacelike
surface to one defined on another:
\begin{eqnarray}
&&\Psi\left[A^{(\sigma'')},\sigma''\right)
\nonumber\\
&&=\int{\cal D}^4\!A^{(\sigma')}
{\cal G}\left[A^{(\sigma'')}\sigma''\right|
\left.A^{(\sigma')}\sigma'\right)
\Psi\left[A^{(\sigma')},\sigma'\right).
\label{Psiprop}
\end{eqnarray}
 The class operators defined from (\ref{formform}) for different
choices of initial and final surfaces are the same so long as all the
spacetime points $\{x\}$ at which the coarse grainings restrict the
fields $A(x)$ still lie in between the initial and final surfaces.

    The conditions satisfied by the wave functional are a consequence
of the gauge invariance of the path integral for the propagator, as
discussed in \cite{JJHJBH}.  Defining co\"{o}rdinates $\{\xi^i\}$ on
the 3-surface $\sigma$ and specifying its embedding in the flat
Minkowski space as $\bbox\{x^\mu\{\xi^i\}\bbox\}$, the metric on the
surface will be
\begin{eqnarray}
ds^2&=&\eta_{\mu\nu}{\partial x^\mu\over\partial\xi^i}d\xi^i
{\partial x^\nu\over\partial\xi^j}d\xi^j=h_{ij}d\xi^i d\xi^j,
\label{metric}
\\
h_{ij}&=&\eta_{\mu\nu}{\partial x^\mu\over\partial\xi^i}
{\partial x^\nu\over\partial\xi^j}.\eqnum{\protect{\ref{metric}}a}
\end{eqnarray}
The condition that $\sigma$ be spacelike means that the three-metric
$\{h_{ij}\}$ is positive definite, so that the volume element on
$\sigma$ is
\begin{equation}
d^3\!\Sigma=d^3\!\xi\sqrt{h},
\end{equation}
where $h=\det\{h_{ij}\}$.  Now the restriction of $A(x)$ onto $\sigma$
is defined by $A^{(\sigma)}(\xi)=A\bbox(x(\xi)\bbox)$, and is a
function of the three co\"{o}rdinates $\{\xi^i\}$ alone.  This is the
first argument of the wave functional $\Psi[A^{(\sigma)},\sigma)$.
Equation (\ref{Psiprop}) shows that the dependence of $\Psi$ on its
first argument is the same as the dependence of the propagator $\cal
G$ on {\em its} first argument.  Since the path integral in
(\ref{formprop}) is invariant under gauge transformations on $A$, the
propagator must be invariant under the effects of those gauge
transformations on $A^{(\sigma')}$ and $A^{(\sigma'')}$.  Since the
gauge transformation
\begin{equation}\eqnum{\protect\ref{trans4A}}
\delta A^a_\mu = -\nabla\!_\mu\,\delta\Lambda_a
- g f_{ab}^c A^c_\mu \delta\Lambda^b
\end{equation}
is nonlocal, the change in $A^{(\sigma)}$ cannot be described by using
only the restriction
$\delta\Lambda^{(\sigma)}(\xi)=\delta\Lambda\bbox(x(\xi)\bbox)$ of the
gauge transformation parameter $\delta\Lambda$ onto the surface
$\sigma$.  To identify the troublesome component of the gradient which
introduces values of $\delta\Lambda$ off of $\sigma$, it is useful to
define a projection tensor
\begin{equation}
\Sigma_\mu^\nu={\partial \xi^i\over\partial x^\mu}
{\partial x^\nu\over\partial\xi^i},
\end{equation}
 where $\partial \xi^i/\partial x^\mu$ is the gradient of $\xi$
with respect to $x$ lying in $\sigma$ so that
\begin{equation}
{\partial \xi^i\over\partial x^\mu}
{\partial x^\mu\over\partial\xi^j}=\delta^i_j.
\end{equation}
 This also follows from the chain rule
\begin{equation}\label{chainrule}
{\partial\over\partial\xi^i}
={\partial x^\nu\over\partial\xi^i}{\partial\over\partial x^\nu}.
\end{equation}
Defining a complementary projection tensor
$\Upsilon_\mu^\nu=\delta_\mu^\nu-\Sigma_\mu^\nu$, we wish to project
out the components of $\{A^\mu\}$ with $\Sigma$ and $\Upsilon$.  Since
$\mathop{\rm Tr}\nolimits\Sigma=3$ and $\mathop{\rm
Tr}\nolimits\Upsilon=1$, it is convenient to define projected objects
with the number of components equal to the rank of the corresponding
projection.  Thus, projections along $\Sigma$ are more concisely
defined by simply projecting with $\partial
x^\nu/\partial\xi^i$.  Defining\footnote{We call this $\aleph_i$
rather than $A^{(\sigma)}_i$ to emphasize that the components
$\aleph_1,\aleph_2,\aleph_3$ are defined with respect to the
co\"{o}rdinates $\xi^1,\xi^2,\xi^3$ lying in the surface $\sigma$ and
are not in general the spatial components
$A^{(\sigma)}_1,A^{(\sigma)}_2,A^{(\sigma)}_3$ defined with respect to
the cartesian spatial co\"{o}rdinates $x^1,x^2,x^3$.}
\begin{equation}
\aleph_i(\xi)={\partial x^\nu\over\partial\xi^i}A^{(\sigma)}_\nu(\xi)
\end{equation}
and using (\ref{chainrule}), the component of (\ref{trans4A}) lying
in $\sigma$ is
\begin{equation}
\delta\aleph^a_i
= -{\partial\delta\Lambda^{(\sigma)}_a\over\partial\xi^i}
- g f_{ab}^c \aleph^c_i \delta\Lambda^{(\sigma)}_b,
\end{equation}
 which is expressed entirely in terms of functions of $\xi$.

    To look at the projection of a four-vector by the rank-one
$\Upsilon$, it is convenient to convert it into a scalar by dotting it
into some arbitrary timelike vector $v$.  Hence the component of $A$
out of the surface $\sigma$ is
\begin{equation}
\phi(\xi)=-v^\mu\Upsilon_\mu^\nu(\xi) A_\nu\bbox(x(\xi)\bbox)
=-u^\nu(\xi) A_\nu\bbox(x(\xi)\bbox).
\end{equation}
 Since $\Upsilon$ has rank one, all possible vectors
$u^\nu=v^\mu\Upsilon_\mu^\nu$ determined from different $v$'s will be
parallel to one another.  Since
\begin{equation}
u^\nu{\partial \xi^i\over\partial x^\nu}=0,
\end{equation}
 $u$ must be parallel to the normal to the surface $\sigma$.  (We
could choose it to be the normal itself, but the normalization factor
will turn out to be irrelevant in what follows.)  Taking the dot
product of (\ref{trans4A}) with $u$, and defining
\begin{equation}
{\partial\over\partial u}=u^\nu{\partial\over\partial x^\nu},
\end{equation}
we have
\begin{equation}
\delta\phi^a
=\left({\partial\delta\Lambda_a\over\partial u}\right)^{(\sigma)}
\!- g f_{ab}^c \phi^c \delta\Lambda^{(\sigma)}_b,
\end{equation}
which cannot be determined from $\phi$ and $\delta\Lambda^{(\sigma)}$
alone.

    Now, since the variation of ${\cal
G}\left[A^{(\sigma)}\sigma\right| \left.A^{(\sigma')}\sigma'\right)$
under a gauge transformation must vanish, this must also be true for
$\Psi$.  That variation is given in terms of the functional
derivatives by
\begin{eqnarray}\label{funk}
\delta\Psi[A^{(\sigma)}]
&&=\int d^3\!\xi\sqrt{h}{{\cal D}\Psi\over{\cal D}A_\mu}\delta A_\mu
\\
=&&\int d^3\!\xi\sqrt{h}
\left(
    {{\cal D}\Psi\over{\cal D}A^{(\sigma)}_\mu}
    \Sigma_\mu^\nu\delta A^{(\sigma)}_\nu
    +{{\cal D}\Psi\over{\cal D}A^{(\sigma)}_\mu}
    \Upsilon_\mu^\nu\delta A^{(\sigma)}_\nu
\right).
\nonumber
\end{eqnarray}
In general, $\Sigma$ and $\Upsilon$ will depend on co\"{o}rdinate
$\xi$, but they will still commute with the gauge transformation
$\delta$ and the functional differentiation $\cal D$.  Put otherwise,
the same amount of information is included in $(\{\aleph_i\},\phi)$ as
in $\left\{A^{(\sigma)}_\mu\right\}
\equiv\left({\bf A}^{(\sigma)},\varphi^{(\sigma)}\right)$, so that
$\Psi$ may be viewed as a functional of $\aleph$ and $\phi$, in which
case (\ref{funk}) becomes
\widetext
\begin{eqnarray}
\delta\Psi[\aleph,\phi]
&=&\int d^3\!\xi\sqrt{h}
\left(
    {{\cal D}\Psi\over{\cal D}\aleph_i}\delta\aleph_i
    +{{\cal D}\Psi\over{\cal D}\phi}\delta\phi
\right)
\nonumber\\
&=&\int d^3\!\xi\sqrt{h}
\Biggl[
    \delta\Lambda^{(\sigma)}_a g f_{ab}^c
    \left(
        \aleph^c_i{{\cal D}\Psi\over{\cal D}\aleph^b_i}
        +\phi^c{{\cal D}\Psi\over{\cal D}\phi^b}
    \right)
    -{\partial\delta\Lambda^{(\sigma)}_a\over\partial\xi^i}
    {{\cal D}\Psi\over{\cal D}\aleph^a_i}
    +\left({\partial\delta\Lambda_a\over\partial u}\right)^{(\sigma)}
    {{\cal D}\Psi\over{\cal D}\phi^a}
\Biggr].
\end{eqnarray}
 Integrating the second term by parts and discarding the term at
spatial infinity gives
\begin{equation}
\delta\Psi[\aleph,\phi]
=\int d^3\!\xi\sqrt{h}
\Biggl\{
    \delta\Lambda^{(\sigma)}_a
    \left[
        gf_{ab}^c\aleph^c_i{{\cal D}\Psi\over{\cal D}\aleph^b_i}
        +gf_{ab}^c\phi^c{{\cal D}\Psi\over{\cal D}\phi^b}
        +{1\over\sqrt{h}}{\partial\over\partial\xi^i}
        \left(\sqrt{h}{{\cal D}\Psi\over{\cal D}\aleph^a_i}\right)
    \right]
    +\left({\partial\delta\Lambda_a\over\partial u}\right)^{(\sigma)}
    {{\cal D}\Psi\over{\cal D}\phi^a}
\Biggr\}.
\end{equation}
\narrowtext
    For this to vanish for arbitrary $\delta\Lambda(x)$, the
coefficients of $\delta\Lambda^{(\sigma)}(\xi)$ and
$(\partial_u\delta\Lambda)^{(\sigma)}(\xi)$ must vanish separately.
This leads to the generalization of (\ref{Psicons}):
\begin{mathletters}
\begin{eqnarray}
{{\cal D}\over{\cal D}\phi^a}\Psi[\aleph,\phi,\sigma)&=&0\\
\left(
    \delta_{ab}{1\over\sqrt{h}}{\partial\over\partial\xi^i}\sqrt{h}
    +gf_{ab}^c\aleph^c_i
\right)
{{\cal D}\over{\cal D}\aleph^b_i}\Psi[\aleph,\phi,\sigma)&=&0.
\end{eqnarray}
\end{mathletters}
 Recognizing the form of the {\em geometric} ``covariant divergence''
on a curved manifold, we see that the general conditions are
\begin{mathletters}\label{Psisigma}
\begin{eqnarray}
{{\cal D}\over{\cal D}\phi}\Psi[\aleph,\psi,\sigma)&=&0\\
D_i{{\cal D}\over{\cal D}\aleph_i}\Psi[\aleph,\psi,\sigma)&=&0,
\end{eqnarray}
\end{mathletters}
 where $D_i$ is the ``covariant'' gradient in both the gauge and
geometric senses of the word:
\begin{eqnarray}
(D_i\zeta^j)_a&=&{\partial\zeta^j_a\over\partial\xi^i}
+\Gamma^j_{ik}\zeta^k_a+gf_{ab}^c\aleph_i^c\zeta^j_b,\label{covgrad}
\\
\Gamma^j_{ik}&=&{h^{j\ell}\over 2}
\left(
    {\partial h_{\ell i}\over\partial\xi^k}
    +{\partial h_{k\ell}\over\partial\xi^i}
    -{\partial h_{ik}\over\partial\xi^\ell}
\right).\eqnum{\protect{\ref{covgrad}}a}
\end{eqnarray}
For the case of $\sigma$ a surface of constant time, (\ref{Psisigma})
reduces to (\ref{Psicons}).

    So, {\em if the initial and final ``times'' are generalized to
arbitrary spacelike surfaces, the conditions (\ref{Psisigma}) obeyed
by the initial and final wave functionals do not truly break Lorentz
invariance,} since {\em they depend only on the surfaces on which the
states are attached, and not on any absolute time direction.} Thus
{\em the entire theory can be formulated in a manifestly Lorentz
invariant way,} at least formally.  With arbitrary initial and final
surfaces, any lattice realization of the path integrals in
(\ref{formform}) will in general involve a non-cartesian lattice.
There are doubtless difficulties in defining such integrals, but they
are beyond the scope of the present work.

 \section{Conclusions}

    In this work, we have developed and examined the
sum-over-histories formulation of generalized quantum mechanics for a
nonabelian gauge theory in the absence of matter, which in addition to
its inherent interest can be viewed as a toy model for Einstein's
general relativity.  The path integrals have been explicitly defined
via an infinitesimal lattice, and shown to be gauge invariant.

    The most general form of the theory allows any set of gauge
invariant phase space alternatives to be assigned a decoherence
functional.  Restricting the alternatives to the phase space
implementations of the gauge electric and magnetic fields and the
covariant derivative gives the ``physical phase space formulation''.
If instead only gauge invariant configuration space alternatives are
considered, we obtain a different subset of possible coarse grainings.
This theory is formally Lorentz-invariant as well.  A further
restriction to coarse grainings involving the configuration space
implementations of gauge electric and magnetic fields and covariant
derivative gives the ``physical configuration space formulation''.

    We have shown that the physical phase space formulation agrees
with a reduced phase space canonical operator (or, as it is known in
other works including \cite{leshouches}, ``ADM'') formulation, so long
as the coarse grainings did not involve time derivatives.  In
particular, the nonabelian Gauss's law constraint ${\bf
D}\cdot\bbox\pi=0$ is always satisfied.

    The physical configuration space formulation behaves slightly
differently. One formally defined quantity which roughly corresponds
to the longitudinal electric field $E_L$ from E\&{}M was shown to
behave in the same way as $E_L$ did in the abelian theory.  I.e.,
coarse grainings by this quantity which decohere predict that it
vanishes.  However, if one coarse grains by more complicated
quantities related to the configuration space constraint $-\bf D\cdot
E$, that may not be so.  In E\&{}M, we have explicitly shown that for
suitable initial conditions, coarse grainings by one such quantity
\{$g[{\bf B}(t_i)]\bigl|\langle E_L\rangle\bigr|^2$\} decohere and
predict non-zero probabilities for the quantity not to vanish.

    Despite the disagreement between the physical configuration space
implementation and reduced phase space operator quantization, the
sum-over-histories formulation is still attractive, since it can be
expressed in a manifestly Lorentz-invariant form.  On the other hand,
the operator theory gives special consideration to the time direction
by singling out the constraint, which is just the timelike component
of the equations of motion $D_\mu G^{\mu\nu}=0$, to be identically
satisfied.

 Since the disagreement between the sum-over-histories theory and a
natural extension of the operator theory comes about when the coarse
graining involves quantities averaged over a spacetime region, as
opposed to the usual quantum mechanical expressions involving
alternatives defined at a single moment of time, perhaps the
sum-over-histories and reduced phase space methods should be seen as
different generalizations of the previously tested formulations (in
which the quantity considered here is not accessible). The Lorentz
invariance of the sum-over-histories method then makes it
the preferred generalization in light of the potential application to
quantum gravity, as it takes one more step towards eliminating the
special role of time in the theory.

    There is also some question as to whether one could construct a
physical apparatus to measure the involved quantity by which we coarse
grained in section~\ref{SSEC:EMK}; on a practical level, the fields
are not directly measurable, but only accessible through their
interactions with charged particles.  It is conceivable that the
differences between the sum-over-histories and operator formalisms are
undetectable in their application to QED and QCD.  However, it is
reasonable to expect that the issues raised by the discrepancy between
them will be relevant to a quantization of GR.  Is enforcement of the
constraints more fundamental than manifest diffeomorphism (here
Lorentz) invariance, or should we only expect the constraints to be
satisfied when the class of alternatives considered singles out the
corresponding time direction in its choice of surfaces?

\acknowledgements

     The author wishes to thank D.~A.~Craig for extensive comments,
and especially J.~B.~Hartle for advice, direction, and encouragement.
He would also like to thank the T-6 group at Los Alamos National
Laboratory, where some of this work was done as a visiting graduate
student.  Some calculations in section~\ref{SSEC:EMK} were worked out
with J.~B.~Hartle.  This material is based upon work supported under a
National Science Foundation Graduate Research Fellowship.  This work
was also supported by NSF grant PHY90-08502.

\appendix

\section{Calculation of class operator for
section~\protect\ref{SSSEC:CONFIGK}}\label{app:NAK}

In this appendix we calculate
\begin{equation}\eqnum{\protect\ref{cD}}
{\cal C}_\Delta=
\left\{
     \prod_{M=J}^0\int{\cal D}{\cal E}^M\exp
     \left[
          i\delta t\int d^3\!x{1\over 2}\left({\cal E}^M\right)^2
     \right]
\right\}
e_\Delta[{\cal E}],
\end{equation}
where
\begin{equation}\eqnum{\protect\ref{eD}}
e_\Delta[{\cal E}]
=\int\limits_\Delta d^{2n}\!f\delta(f-\langle{\cal E}\rangle)
\end{equation}
 describes the coarse graining by values of the mode average
\begin{equation}\eqnum{\protect\ref{modeave}}
\langle{\cal E}\rangle
={1\over\Omega}\int\limits_\Omega dt\,d^3\!k\,
{\cal E}_{\bf k}(t)
={1\over\Omega}\sum_{M\in\Omega}\delta t\int\limits_\Omega d^3\!k\,
{\cal E}^M_{\bf k}.
\end{equation}
While ${\cal E}({\bf x})$ is a real quantity, the Fourier transform
\begin{equation}\eqnum{\protect\ref{fourE}}
{\cal E}^M_{\bf k}=\int{d^3\!x\over(2\pi)^{3/2}}
e^{-i\bbox{\rm k}\cdot\bbox{\rm x}}{\cal E}^M({\bf x})
\end{equation}
is complex but constrained to obey ${\cal E}^*_{\bf k}={\cal E}_{-\bf
k}$.  Integrating over the independent degrees of freedom in Fourier
space necessitates the development of more notation.  Letting a
superscript of R or I indicate the real or imaginary part,
respectively, of a complex number, and using the Jacobian determinant
calculated in appendix~\ref{app:jacob} for the discrete Fourier
transform, the path integral measure is (using the infinite numerical
constant $\Xi$ defined in appendix~\ref{app:jacob})
\begin{eqnarray}\label{measE}
{\cal D}{\cal E}^M&=&\prod_{a,\bf x}N_A d{\cal E}^M_a({\bf x})
\nonumber\\
&=&\prod_{a,\bf x}N_A
d{\cal E}^{M\rm R}_a({\bf x})d{\cal E}^{M\rm I}_a({\bf x})
\delta\biglb({\cal E}^{M\rm I}_a({\bf x})\bigrb)
\nonumber\\
&=&\prod_a
\left(
     \prod_{\bf k}N_A {\delta^3\!k\over\delta^3\!x}
     d{\cal E}^{M\rm R}_{a,\bf k} d{\cal E}^{M\rm I}_{a,\bf k}
\right)
\\
&&\times\Xi
{\prod_{\bf k}}^{1/2}{\delta^3\!x\over\delta^3\!k}
\delta\left({\cal E}^{M\rm R}_{a,\bf k}
            -{\cal E}^{M\rm R}_{a,-\bf k}\right)
\delta\left({\cal E}^{M\rm I}_{a,\bf k}
            +{\cal E}^{M\rm I}_{a,-\bf k}\right),
\nonumber
\end{eqnarray}
where the $\prod^{1/2}$ means we are only taking the product over half
the modes (leaving out the redundant ones, whose spatial frequency is
minus the spatial frequency of a mode already counted), so that
$\prod_{\bf x}f^{\rm R}({\bf x})$ and $\prod_{\bf k}^{1/2}f^{\rm
R}_{\bf k}f^{\rm I}_{\bf k}$ each have the same number of factors.
The factor in $\prod^{1/2}$ for the zero mode\footnote{There will
typically also be modes on the boundary of
spatial frequency space which are identified with the corresponding
modes on the opposite boundary, and so that for these ${\bf k}$'s
${\cal E_{\bf k}}={\cal E_{-\bf k}}$ as with the zero mode ${\bf
k}={\bf 0}$.  For example, in the discrete Fourier transform on a
one-dimensional lattice \cite{numrec} with an even number $N$ of
points, the modes of frequency $1/2N$ and $-1/2N$ (the Nyquist
critical frequency and its image) are identified, so the situation is
analogous to that of the zero mode.  The identification, when combined
with the condition ${\cal E}^*_{\bf k}={\cal E}_{-\bf k}$, requires
that the Fourier components on the boundary be real. The boundary is
not a region of interest to us in spatial frequency space, and we
assume that the prescription for those factors is similar to the one
for the zero mode.\label{fn:Nyquist}} is understood to be
\addtocounter{equation}{-1}
\begin{mathletters}
\begin{equation}
 \left({\delta^3\!x\over\delta^3\!k}\right)^{1/2}
\delta({\cal E}^{M\rm I}_{a,\bf 0}).
\end{equation}
\end{mathletters}
We can use the delta functions to perform the integrals over half of
the Fourier components so that
\begin{eqnarray}
{\cal D}{\cal E}^M&=&\prod_a\Xi
{\prod_{\bf k}}^{1/2}
N_A^2{\delta^3\!k\over\delta^3\!x}
d{\cal E}^{M\rm R}_{a,\bf k} d{\cal E}^{M\rm I}_{a,\bf k}
\nonumber\\
&=&\Xi^n
{\prod_{\bf k}}^{1/2}
\left(N_A^2{\delta^3\!k\over\delta^3\!x}\right)^n
d^n\!{\cal E}^{M\rm R}_{\bf k} d^n\!{\cal E}^{M\rm I}_{\bf k}
\end{eqnarray}
with the factor for the zero mode understood to be
\addtocounter{equation}{-1}
\begin{mathletters}
\begin{equation}
\left(N_A^2{\delta^3\!k\over\delta^3\!x}\right)^{n/2}
d^n\!{\cal E}^{M\rm R}_{\bf 0}.
\end{equation}
\end{mathletters}

	The relevant part of the Lagrangian is
\begin{eqnarray}
{1\over 2}\int d^3\!x\left[{\cal E}^M({\bf x})\right]^2&=&
{1\over 2}\int d^3\!x\left|{\cal E}^M({\bf x})\right|^2=
{1\over 2}\int d^3\!k\left|{\cal E}^M_{\bf k}\right|^2
\nonumber\\
&=&
{\sum_{\bf k}}^{1/2}\delta^3\!k\left|{\cal E}^M_{\bf k}\right|^2;
\end{eqnarray}
defining $\omega=\delta t\delta^3\!k$, we have
\begin{eqnarray}
{\cal C}_\Delta&=&
\Bigglb(
     \prod_{M=J}^0\Xi^n{\prod_{\bf k}}^{1/2}
     \left(N_A^2 {\delta^3\!k\over\delta^3\!x}\right)^n
\nonumber\\
     &&\times
     \int
     d^n\!{\cal E}^{M\rm R}_{\bf k}d^n\!{\cal E}^{M\rm I}_{\bf k}\exp
     \left\{
          i\omega
          \left[
               \left({\cal E}^{M\rm R}_{\bf k}\right)^2
               +\left({\cal E}^{M\rm I}_{\bf k}\right)^2
          \right]
     \right\}
\Biggrb)
\nonumber\\
&&\times
e_\Delta[{\cal E}].
\end{eqnarray}
If we use $\lambda$ as a mode label, combining $M$ and $\bf k$,
\begin{eqnarray}
&&e_\Delta[{\cal E}]
\nonumber\\
&&=\int\limits_\Delta d^n\!f^{\rm R}d^n\!f^{\rm I}
\delta^n\!
\biggl(
     f^{\rm R}
     -{\omega\over\Omega}
      \sum_{\lambda\in\Omega}{\cal E}^{\rm R}_\lambda
\biggr)
\delta^n\!
\biggl(
     f^{\rm I}
     -{\omega\over\Omega}
      \sum_{\lambda\in\Omega}{\cal E}^{\rm I}_\lambda
\biggr)
\nonumber\\
&&=\int\limits_\Delta d^{2n}\!f
\delta^{2n}\!
\biggl(
     f-{\omega\over\Omega}\sum_{\lambda\in\Omega}{\cal E}_\lambda
\biggr)
\end{eqnarray}
and
\begin{eqnarray}\label{CD}
{\cal C}_\Delta&=&
\Bigglb(
     \Xi^{n(J+1)}
     {\prod_\lambda}
     \left(N_A^2 {\delta^3\!x\over\delta^3\!k}\right)^n
\\
     &&\,\,\times
     \int
     d^n\!{\cal E}^{\rm R}_\lambda d^n\!{\cal E}^{\rm I}_\lambda\exp
     \left\{
          i\omega
          \left[
               \left({\cal E}^{\rm R}_\lambda\right)^2
               +\left({\cal E}^{\rm I}_\lambda\right)^2
          \right]
     \right\}
\Biggrb)
e_\Delta[{\cal E}].
\nonumber
\end{eqnarray}
We can factor the product in (\ref{CD}) into a product over modes in
$\Omega$ and one over modes not in $\Omega$.  The latter is a constant
which is the same for all alternatives $\{c_\Delta\}$:
\begin{eqnarray}
{\cal K}=
\Bigglb(&&
     \Xi^{n(J+1)}
     {\prod_{\lambda\notin\Omega}}
     \left(N_A^2 {\delta^3\!x\over\delta^3\!k}\right)^n
\nonumber\\
     &&\times
     \int
     d^n\!{\cal E}^{\rm R}_\lambda d^n\!{\cal E}^{\rm I}_\lambda\exp
     \left\{
          i\omega
          \left[
               \left({\cal E}^{\rm R}_\lambda\right)^2
               +\left({\cal E}^{\rm I}_\lambda\right)^2
          \right]
     \right\}
\Biggrb),
\end{eqnarray}
which leaves
\begin{eqnarray}
{\cal C}_\Delta={\cal K}&&
\Bigglb(
     \prod_{\lambda\in\Omega}
     \left(N_A^2 {\delta^3\!x\over\delta^3\!k}\right)^n
\nonumber\\
     &&\quad\times
     \int
     d^n\!{\cal E}^{\rm R}_\lambda d^n\!{\cal E}^{\rm I}_\lambda\exp
     \left\{
          i\omega
          \left[
               \left({\cal E}^{\rm R}_\lambda\right)^2
               +\left({\cal E}^{\rm I}_\lambda\right)^2
          \right]
     \right\}
\Biggrb)
\nonumber\\
&&\times
\int\limits_\Delta d^{2n}\!f\,
\delta^{2n}\!
\biggl(
     f-{\omega\over\Omega}\sum_{\lambda\in\Omega}{\cal E}_\lambda
\biggr).
\end{eqnarray}
If we define $N=\sum_{\lambda\in\Omega}1$ to be the number of modes
in $\Omega$, and write all the modes of ${\cal E}$ in $\Omega$ as a
$2N$-component column vector:
\begin{equation}
{\cal E}=
\left(
    \begin{array}{c}
          {\cal E}^{\rm R}_\lambda\\
          {\cal E}^{\rm I}_\lambda
     \end{array}
\right),
\end{equation}
we have
\begin{eqnarray}
{\cal C}_\Delta&=&{\cal K}
\left[
     \prod_{\lambda\in\Omega}
     \left(N_A^2 {\delta^3\!x\over\delta^3\!k}\right)^n \int
     d^n\!{\cal E}^{\rm R}_\lambda d^n\!{\cal E}^{\rm I}_\lambda
\right]
e^{i\omega{\cal E}^2}
\nonumber\\
&&\times
\int\limits_\Delta d^{2n}\!f\,
\delta^{2n}\!
\biggl(
     f-{\omega\over\Omega}\sum_{\lambda\in\Omega}{\cal E}_\lambda
\biggr).
\end{eqnarray}
We define a column vector $\Upsilon$ to be the discrete Fourier
transform of $\cal E$:
\begin{eqnarray}
\Upsilon&=&
\left(
    \begin{array}{c}
          \Upsilon^{\rm R}_{w\bf y}\\
          \Upsilon^{\rm I}_{w\bf y}
     \end{array}
\right)
\nonumber\\
&=&
\left(
    \begin{array}{cc}
          {1\over 2}       & {1\over 2} \\
          {1\over 2i}    & -{1\over 2i}
     \end{array}
\right)
\left(
    \begin{array}{cc}
     {\omega\over\Omega} e^{i\bbox{\rm y}\cdot\bbox{\rm k}+iwt}  & 0 \\
     0 &  {\omega\over\Omega} e^{-i\bbox{\rm y}\cdot\bbox{\rm k}-iwt}\\
    \end{array}
\right)
\nonumber\\
&&\times
\left(
    \begin{array}{cc}
          1    & i \\
          1    & -i
     \end{array}
\right)
\left(
    \begin{array}{c}
          {\cal E}^{\rm R}_{t\bf k}\\
          {\cal E}^{\rm I}_{t\bf k}
     \end{array}
\right)
=M{\cal E}.\label{fourY}
\end{eqnarray}
 [This is a rigorous version of the traditional treatment of the
complex $\Upsilon$ and $\Upsilon^*$ as independent variables; the
middle matrix of the product of three is the one which would be used
to convert the column vector
 $\left({{\cal E}_{t\bf k}\atop {\cal E}^*_{t\bf k}}\right)$ to
 $\left({\Upsilon_{w\bf y}\atop\Upsilon^*_{w\bf y}}\right)$.] The zero
components of $\Upsilon$ are
\begin{equation}
\Upsilon_{0\bf 0}
={\omega\over\Omega}\sum_{\lambda\in\Omega}{\cal E}_\lambda
=\langle{\cal E}\rangle.
\end{equation}

     Since $M$ is a real matrix which is the product of three
matrices, each of which is proportional to a unitary matrix, it must
be proportional to an orthogonal (i.e., real and unitary) matrix $\cal
M$.  The calculation in appendix~\ref{app:jacob} yields
\begin{equation}
\det M=N^{-N}=(\det{\cal M})(N^{-1/2})^{2N},
\end{equation}
so we have $M={\cal M}/\sqrt{N}$.  Thus
\begin{equation}
\Upsilon^2={\cal E}^{\rm TR}M^{\rm TR}M{\cal E}=
{{\cal E}^{\rm TR}{\cal M}^{\rm TR}{\cal ME}\over N}
={{\cal E}^2\over N},
\end{equation}
so $\omega{\cal E}^2=N\omega\Upsilon^2=\Omega\Upsilon^2$ and
\widetext
\begin{eqnarray}
{\cal C}_\Delta&=&{\cal K}
\left[
     \prod_{w\bf y}
     \left(N_A^2 {\delta^3\!x\over\delta^3\!k}\right)^n \int N
     d^n\!\Upsilon^{\rm R}_{w\bf y} d^n\!\Upsilon^{\rm I}_{w\bf y}
\right]
e^{i\Omega\Upsilon^2}
\int\limits_\Delta d^{2n}\!f\,
\delta^{2n}(f-\Upsilon_{0\bf 0})\nonumber\\
&=&{\cal K}
\Bigglb(
     \prod_{w\bf y}
     \left(N_A^2 {\delta^3\!x\over\delta^3\!k}\right)^n \int N
     d^n\!\Upsilon^{\rm R}_{w\bf y} d^n\!\Upsilon^{\rm I}_{w\bf y}\exp
     \left\{
          i\Omega
          \left[
               \left(\Upsilon^{\rm R}_{w\bf y}\right)^2
               +\left(\Upsilon^{\rm I}_{w\bf y}\right)^2
          \right]
     \right\}
\Biggrb)
\int\limits_\Delta d^{2n}\!f\,
\delta^{2n}(f-\Upsilon_{0\bf 0}).
\end{eqnarray}
\narrowtext

     Factoring all the $\{\Upsilon_{w\bf y}\}$ except the zero mode
into the constant, and using the delta function to do the
$\Upsilon_{0\bf 0}$ integrals, we have
\begin{equation}\label{eqA}
{\cal C}_\Delta
={\cal K}'\int\limits_\Delta d^{2n}\!f\, e^{i\Omega|f|^2}.
\end{equation}

\section{Calculation of Jacobian determinants for discrete Fourier
transforms}\label{app:jacob}

Given a complex function $f(x)$ of a $D$ dimensional variable
$x=\{x_\alpha|\alpha=1,2,\ldots D\}$, if we define $f$ only on a
spatial lattice with $N_\alpha$ lattice points in the $\alpha$
direction (and thus $\prod_{\alpha=1}^D N_\alpha\equiv N$ total
lattice points), we have a vector
\begin{equation}
f=\left(\begin{array}{c}f^{\rm R}_x\\f^{\rm I}_x\end{array}\right)
\end{equation}
with $2N$ real components.  If we take the Fourier transform (see
\cite{numrec} for a general treatment of the discrete Fourier
transform)
\begin{equation}
F_k=\sum_x e^{\mp ik\cdot x} f_x,
\end{equation}
there is a corresponding matrix transformation\footnote{As with
(\protect\ref{fourY}), this treatment is the more careful analog of
treating $f_x$ and $f_x^*$ as independent variables.} on
${\cal R}^{2N}$:
\begin{eqnarray}
\left(\begin{array}{c}F^{\rm R}_k\\F^{\rm I}_k\end{array}\right)&=&
\left(
	\begin{array}{cc}{1\over 2}&{1\over 2}\\
	{1\over 2i}&-{1\over 2i}\end{array}
\right)
\nonumber\\
&&\times
\left(
	\begin{array}{cc}\exp(\mp i\sum_\alpha k_\alpha x_\alpha)&0\\
	0&\exp(\pm i\sum_\alpha k_\alpha x_\alpha)\end{array}
\right)
\nonumber\\
&&\times
\left(\begin{array}{cc}1&i\\1&-i\end{array}\right)
\left(\begin{array}{c}f^{\rm R}_x\\f^{\rm I}_x\end{array}\right)
\label{FMf}
\end{eqnarray}
or
\begin{equation}
F=Mf.\eqnum{\protect{\ref{FMf}}$'$}
\end{equation}
The Jacobian of this transformation is given by $\det M$.  Since the
first and third of the three matrices of which $M$ is a product are
inverses of each other and the second is block diagonal, we have
\begin{eqnarray}
\det_{2N\times 2N}M
&=&\det_{N\times N}e^{\mp ik\cdot x}\det_{N\times N}e^{\pm ik\cdot x}
\nonumber\\
&=&\det_{N\times N}
\left(\sum_k e^{\mp ix\cdot k}e^{\pm ik\cdot y}\right)
\nonumber\\
&=&\det_{N\times N}(N\delta_{xy})=N^N.
\end{eqnarray}

To apply this to the transformations in section \ref{SSEC:CONFIGK}, we
need to take into account the normalization constants.  In the
discrete case, (\ref{fourE}) becomes
\begin{equation}
{\cal E}_{a,\bf k}^M=\sum_{\bf x}{\delta^3\!x\over(2\pi)^{3/2}}
e^{-i\bbox{\rm k}\cdot\bbox{\rm x}}{\cal E}_a^M({\bf x})
\end{equation}
and the Jacobian is
\begin{equation}\label{jacN}
N^N\left({\delta^3\!x\over(2\pi)^{3/2}}\right)^{2N}.
\end{equation}
 Now, the relationship between $\delta^3\!x$, the number of spatial
lattice sites $N$, and the lattice spacing $\delta^3\!k$ in spatial
frequency can be deduced by geometric arguments, but the simplest
method is to note that
\begin{equation}
{\cal E}_a^M({\bf x})=\sum_{\bf k}{\delta^3\!k\over(2\pi)^{3/2}}
e^{i\bbox{\rm x}\cdot\bbox{\rm k}}{\cal E}_{a,\bf k}^M
\end{equation}
and hence
\begin{eqnarray}
\left(
	\begin{array}{c}{\cal E}_a^{M\rm R}({\bf x})\\
	{\cal E}_a^{M\rm I}({\bf x})\end{array}
\right)&=&
\left(
	\begin{array}{cc}{1\over 2}&{1\over 2}\\
	{1\over 2i}&-{1\over 2i}\end{array}
\right)
\nonumber\\
&&\times
\left(
     \begin{array}{cc}
 {\delta^3\!k\over(2\pi)^{3/2}}e^{i\bbox{\rm x}\cdot\bbox{\rm k}} & 0\\
 0 & {\delta^3\!k\over(2\pi)^{3/2}}e^{-i\bbox{\rm x}\cdot\bbox{\rm k}}
     \end{array}
\right)
\nonumber\\
&&\times
\left(
     \begin{array}{cc}
{\delta^3\!x\over(2\pi)^{3/2}}e^{-i\bbox{\rm k}\cdot\bbox{\rm y}} & 0\\
0 &d{\delta^3\!x\over(2\pi)^{3/2}}e^{i\bbox{\rm k}\cdot\bbox{\rm y}}
     \end{array}
\right)
\nonumber\\
&&\times
\left(\begin{array}{cc}1&i\\1&-i\end{array}\right)
\left(
	\begin{array}{c}{\cal E}_a^{M\rm R}({\bf y})\\
	{\cal E}_a^{M\rm I}({\bf y})\end{array}
\right).
\end{eqnarray}
Taking the determinant, we find
\begin{equation}
1=\left({\delta^3\!k\over(2\pi)^{3/2}}\right)^{2N}N^N
\left({\delta^3\!x\over(2\pi)^{3/2}}\right)^{2N}N^N
\end{equation}
or
\begin{equation}
N={(2\pi)^3\over\delta^3\!k\delta^3\!x}.
\end{equation}
Substituting into (\ref{jacN}), we see that the Jacobian is
\begin{equation}
\left({(2\pi)^3\over\delta^3\!k\delta^3\!x}\right)^N
\left({\delta^3\!x\over(2\pi)^{3/2}}\right)^{2N}
=\left(\delta^3\!x\over\delta^3\!k\right)^N
\end{equation}
so that
\begin{eqnarray}
\prod_{\bf x}d{\cal E}_a^{M\rm R}({\bf x})
d{\cal E}_a^{M\rm I}({\bf x})
&=&\left(\delta^3\!k\over\delta^3\!x\right)^N
\prod_{\bf k}d{\cal E}_{a,\bf k}^{M\rm R}
d{\cal E}_{a,\bf k}^{M\rm I}
\nonumber\\
&=&\prod_{\bf k}\left(\delta^3\!k\over\delta^3\!x\right)
d{\cal E}_{a,\bf k}^{M\rm R}d{\cal E}_{a,\bf k}^{M\rm I},
\end{eqnarray}
which is the correct factor for (\ref{measE}).

Equation (\ref{measE}) also involves the Jacobian for the
transformation of the delta functions
\begin{equation}
\prod_{\bf x}\delta\biglb({\cal E}^{M\rm I}_a({\bf x})\bigrb)
\end{equation}
into
\begin{equation}
{\prod_{\bf k}}^{1/2}
\delta\left({\cal E}^{M\rm R}_{a,\bf k}
            -{\cal E}^{M\rm R}_{a,-\bf k}\right)
\delta\left({\cal E}^{M\rm I}_{a,\bf k}
            +{\cal E}^{M\rm I}_{a,-\bf k}\right).
\end{equation}
To determine that, define $F_k^\pm=F_k\pm F_{-k}$ and observe that
\begin{eqnarray}
\prod_x df^{\rm R}_x df^{\rm I}_x &=&
N^N \prod_k dF^{\rm R}_k dF^{\rm I}_k
\nonumber\\
&=& N^N \prod_k{}^{1/2} dF^{\rm R}_k dF^{\rm R}_{-k}
dF^{\rm I}_k dF^{\rm I}_{-k}
\nonumber\\
&=& N^N \prod_k{}^{1/2}
{dF^{{\rm R}+}_k dF^{{\rm R}-}_k\over 2}
{dF^{{\rm I}+}_k dF^{{\rm I}-}_k\over 2},
\end{eqnarray}
so
\begin{eqnarray}
&&\prod_x \delta(f^{\rm R}_x) \delta(f^{\rm I}_x)
\nonumber\\
&&= N^{-N} \prod_k{}^{1/2}
2 \delta(F^{{\rm R}+}_k) \delta(F^{{\rm R}-}_k)
2 \delta(F^{{\rm I}+}_k) \delta(F^{{\rm I}-}_k).
\end{eqnarray}
We assume by symmetry that when we factor the Jacobian splits evenly:
\begin{mathletters}
\begin{eqnarray}
\prod_x \delta(f^{\rm R}_x) &=& N^{-N/2} \prod_k{}^{1/2}
2 \delta(F^{{\rm R}+}_k) \delta(F^{{\rm I}-}_k)
\\
\prod_x \delta(f^{\rm I}_x) &=& N^{-N/2} \prod_k{}^{1/2}
2 \delta(F^{{\rm R}-}_k) \delta(F^{{\rm I}+}_k);
\end{eqnarray}
\end{mathletters}
this means that
\begin{eqnarray}
&&\prod_{\bf x}\delta\biglb({\cal E}^{M\rm I}_a({\bf x})\bigrb)
\\
&&=
\Xi{\prod_{\bf k}}^{1/2} \left(\delta^3\!x\over\delta^3\!k\right)
\delta\left({\cal E}^{M\rm R}_{a,\bf k}
            -{\cal E}^{M\rm R}_{a,-\bf k}\right)
\delta\left({\cal E}^{M\rm I}_{a,\bf k}
            +{\cal E}^{M\rm I}_{a,-\bf k}\right).
\nonumber
\end{eqnarray}
$\Xi$ is not quite equal to $2^{N/2}$ because the analysis above does
not go through for the zero mode and some modes on the boundary (see
footnote~\ref{fn:Nyquist}, page~\pageref{fn:Nyquist}) which are
identified with their images.  In those cases, the analysis produces
the same Jacobian, only without the factor of $2$.  At any rate, $\Xi$
is a constant, and its precise value is unimportant.

The determinant of the transformation (\ref{fourY}) is even more
straightforward.  There the number of modes is just
$N={\Omega\over\omega}$, and the determinant is thus
\begin{equation}
\left({\omega\over\Omega}\right)^{2N}N^N=N^{-N}.
\end{equation}

\section{Calculation of the decoherence functional for
section~\protect\ref{SSEC:EMK} in the presence of a Gaussian initial
state}\label{app:pb}

    Here we calculate the decoherence functional (\ref{DGmath}) for an
initial state where $p(b)$ [cf.\ (\ref{probb})] is a Gaussian in
$B={1\over b}$:
\begin{equation}
p(B^{-1})=A_\lambda\Theta(B-\lambda)e^{-B^2/2\sigma^2}.
\end{equation}
Then (\ref{Gy}) becomes
\begin{eqnarray}
G(y)&\propto&\int_\lambda^\infty dB\, e^{i\Omega yB}e^{-B^2/2\sigma^2}
\\
&=&\Lambda_1(y)+e^{-(\Omega\sigma y)^2/2}
\int_0^\infty dB e^{-(B-i\Omega\sigma^2 y)^2/2\sigma^2},\label{Geq}
\nonumber
\end{eqnarray}
where
\begin{equation}
\Lambda_1(y)=-\int_0^\lambda dB\, e^{-B^2/2\sigma^2}e^{i\Omega y B}
\end{equation}
satisfies
\begin{equation}
|\Lambda_1(y)|<\lambda.
\end{equation}
 The second term in (\ref{Geq}) can be massaged by deformation of
contour to give
\begin{equation}
G(y)\propto\sqrt{2\over\pi}{\Lambda_1(y)\over\sigma}+G_R(y)+iG_I(y),
\end{equation}
where
\addtocounter{equation}{-1}
\begin{mathletters}
\begin{equation}
G_R(y)=e^{-(\Omega\sigma y)^2/2}
\end{equation}
and
\begin{equation}
G_I(y)=\sqrt{2\over\pi}\Omega\sigma\int_0^y dz\,
e^{(\Omega\sigma)^2(z^2-y^2)/2}.
\end{equation}
\end{mathletters}

    If we choose the bins to be of a uniform size $\Delta$:
\begin{equation}
\Delta_J\equiv[J\Delta,(J+1)\Delta), \quad 0\le J\in{\cal Z},
\end{equation}
we have
\begin{eqnarray}
\mathop{\rm Re}\nolimits D(J,J')&\propto&
\int_{J\Delta}^{(J+1)\Delta}\!df
\int_{J'\Delta}^{(J'+1)\Delta}\!df'
e^{-(\Omega\sigma)^2(f-f')^2/2}
\nonumber\\
&&+\sqrt{2\over\pi}{\Lambda_2(J-J',\Delta)\Delta^2\over\sigma},
\end{eqnarray}
where
\begin{eqnarray}
&&\Lambda_2(J-J',\Delta)
\nonumber\\
&&=\Delta^{-2}\int_{J\Delta}^{(J+1)\Delta}\!df
\int_{J'\Delta}^{(J'+1)\Delta}\!df'
\mathop{\rm Re}\nolimits\Lambda_1(f-f')
\end{eqnarray}
again satisfies
\begin{equation}
|\Lambda_2(J-J',\Delta)|<\lambda.
\end{equation}
 Making a suitable change of variables and factoring out $\Delta^2$,
we obtain
\begin{eqnarray}
\mathop{\rm Re}\nolimits D(J,J')&\propto& D_+(|J-J'|)+D_-(|J-J'|)
\nonumber\\
&&+\sqrt{2\over\pi}{\Lambda_2(J-J',\Delta)\over\sigma},
\label{ReD}
\end{eqnarray}
 where $D_\pm$ is the contribution to the double integral from
$\pm[f-f'-(J-J')\Delta](J-J')>0$:
\addtocounter{equation}{-1}
\begin{mathletters}
\begin{equation}\label{pm}
D_\pm(\Delta{J})=\int_0^1 d\eta\,(1-\eta)
\exp[-(\Omega\sigma\Delta)^2(\Delta{J}\pm\eta)^2/2]
\end{equation}
\end{mathletters}
(Fig.~\ref{fig:pm}).  Now,
\begin{eqnarray}
D_+(\Delta{J})&\le& e^{-(\Omega\sigma\Delta)^2(\Delta{J})^2/2}
\int_0^1 d\eta\,(1-\eta) e^{-(\eta\Omega\sigma\Delta)^2/2}
\nonumber\\
&=&e^{-(\Omega\sigma\Delta)^2(\Delta{J})^2/2}D_+(0).\label{Dp}
\end{eqnarray}
 From the definition (\ref{pm}) it is evident that $D_-(0)=D_+(0)$.
For $\Delta{J}\ge 1$,
\begin{eqnarray}
D_-(\Delta{J})
&=&\int_0^1d\eta\,\eta
\exp\{-(\Omega\sigma\Delta)^2[(\Delta{J}-1)+\eta]^2/2\}
\nonumber\\
&\le& e^{-(\Omega\sigma\Delta)^2(\Delta{J}-1)^2/2}D_-(1).
\label{Dm}
\end{eqnarray}
Combining these results, we see\footnote{Recall that $D(J,J)$ is real
and positive by (\ref{herm}-\ref{pos}).}
\widetext
\begin{equation}\label{blat}
{\mathop{\rm Re}\nolimits D(J+\Delta{J},J)\over D(J,J)}\le
{e^{-(\Omega\sigma\Delta)^2|\Delta{J}|^2/2}D_+(0)
+e^{-(\Omega\sigma\Delta)^2(|\Delta{J}|-1)^2/2}D_-(1)
+\sqrt{2/\pi}\,{\Lambda_2(\Delta{J},\Delta)/\sigma}
\over
2D_+(0)+\sqrt{2/\pi}\,{\Lambda_2(\Delta{J},\Delta)/\sigma}}.
\end{equation}
\narrowtext
So we have reduced the question of whether we have weak decoherence
[$|\mathop{\rm Re}\nolimits D(J+\Delta{J})|\ll D(J,J)$ for
$\Delta{J}\ne 0$] to a calculation of $D_+(0)$ and $D_-(1)$.  It is
straightforward to show
\begin{equation}
D_-(1)={1-e^{-(\Omega\sigma\Delta)^2/2}
\over(\Omega\sigma\Delta)^2}
\end{equation}
and
\begin{equation}
D_+(0)={\sqrt{\pi/2}\over\Omega\sigma\Delta}
\mathop{\rm erf}\nolimits\left({\Omega\sigma\Delta\over\sqrt{2}}\right)
-D_-(1),
\end{equation}
where $\mathop{\rm erf}\nolimits(z)$ is the error function
$\mathop{\rm erf}\nolimits(z)={2\over\sqrt{\pi}}\int_0^z e^{-t^2}
dt$, which satisfies $\mathop{\rm erf}\nolimits(\infty)=1$ and
$\mathop{\rm erf}\nolimits{z}\ge 1-e^{-z^2}$.  Thus
\begin{eqnarray}
D_+(0)&\ge&\sqrt{\pi\over 2}
{1-e^{-(\Omega\sigma\Delta)^2/2}\over\Omega\sigma\Delta}
-{1-e^{-(\Omega\sigma\Delta)^2/2}\over(\Omega\sigma\Delta)^2}
\nonumber\\
&=&\left(1-e^{-(\Omega\sigma\Delta)^2/2}\right)
{\Omega\sigma\Delta\sqrt{\pi/2}-1\over(\Omega\sigma\Delta)^2}.
\end{eqnarray}
This means that if the cutoff $\lambda\lesssim\sigma
e^{-(\Omega\sigma\Delta)^2/2}$,
(\ref{blat}) becomes, to lowest order
in $e^{-(\Omega\sigma\Delta)^2/2}$,
\begin{eqnarray}
&&{\mathop{\rm Re}\nolimits D(J+\Delta{J},J)\over D(J,J)}
\nonumber\\
&\lesssim&
{\exp[-(\Omega\sigma\Delta)^2(|\Delta{J}|-1)^2/2]D_-(1)\over 2D_+(0)}
+O\left({\lambda\over\sigma}\right)
\nonumber\\
&\lesssim&{\exp[-(\Omega\sigma\Delta)^2(|\Delta{J}|-1)^2/2]
\over\Omega\sigma\Delta\sqrt{2\pi}}
+O\left({\lambda\over\sigma}\right).
\end{eqnarray}
For large $\Omega\sigma\Delta$, the $\lambda$-dependent term will not
be relevent to the issue of decoherence.

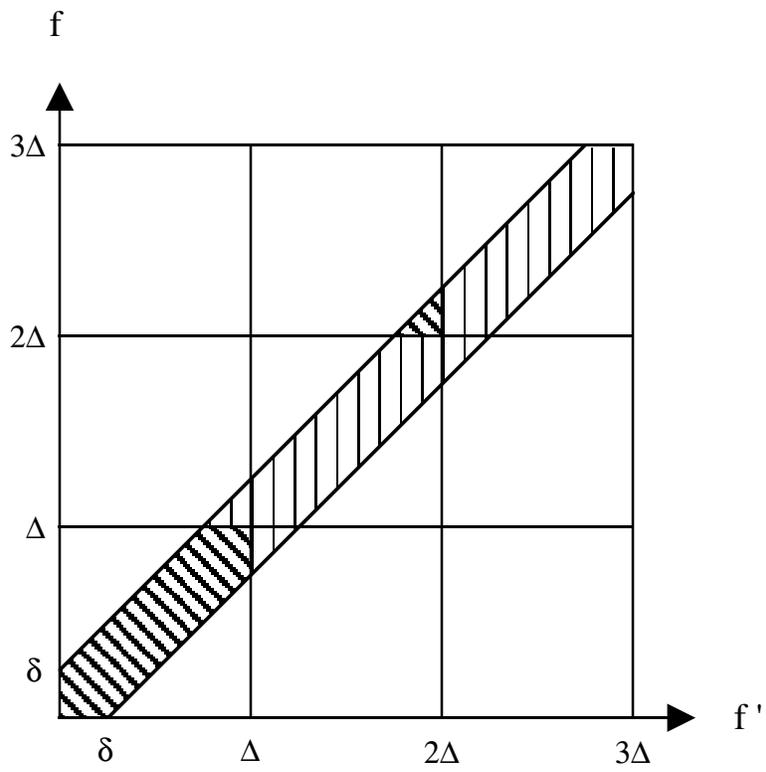
\begin{figure}
\caption{The regions of integration of a general $G(f-f')$ to produce
the decoherence functional $D(J,J')$ via (\protect\ref{DG}).  If
$G(y)$ is negligible for $|y|\protect\gtrsim\delta$, integrals of
$G(y)$ over regions two or more spots off the diagonal ($|J-J'|\ge 2$)
will be negligible.  Squares on the diagonal ($J=J'$) have a region of
area $2\Delta\delta-\delta^2$ over which $G(y)$ is appreciable.
Squares one spot off the diagonal ($|J-J'|=1$) include some
non-negligible values of $G(y)$, but only in a triangular region of
area $\delta^2/2$.  Thus $D(J,J\pm 1)$ should be suppressed by a
factor of $\delta/\Delta$ relative to $D(J,J)$.  Compare Fig.~1 of
\protect\cite{classeq}.}
\label{fig:dropoff}
\end{figure}

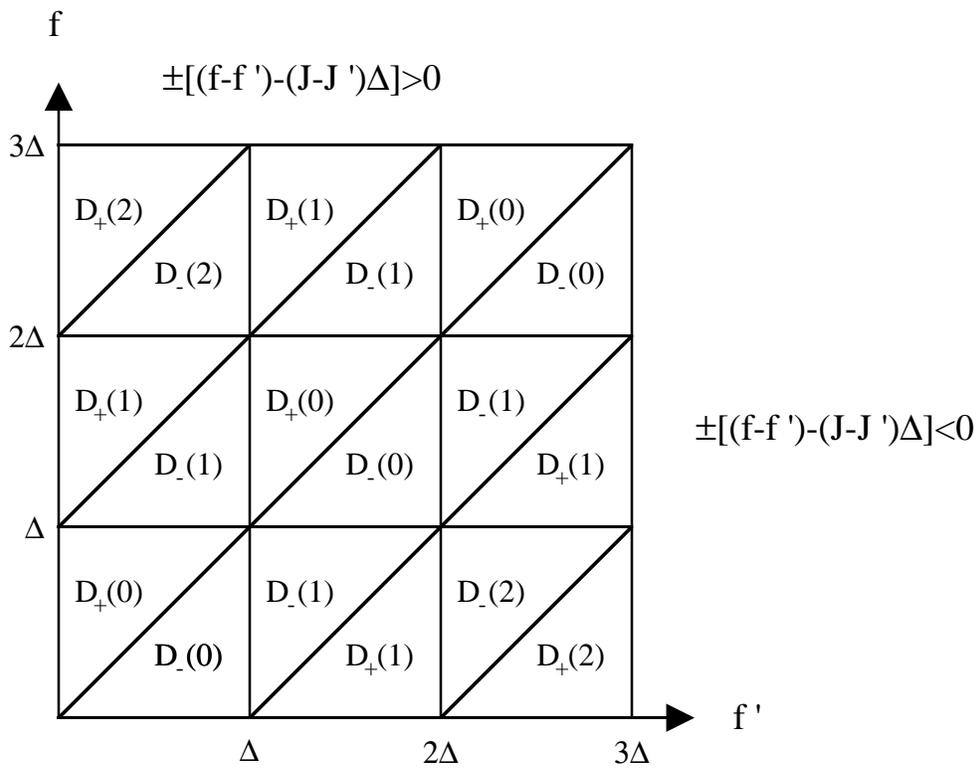
\begin{figure}
\caption{The regions of integration for $D_\pm(|\Delta{J}|)$.  Because
of the exponential dropoff in $\protect{\mathop{\rm Re}\nolimits}
G(f-f')$ as one moves towards larger $|f-f'|$, $D_+(|\Delta{J}|)$ is
reduced from $D_+(0)$ by a factor of
$e^{-(\Omega\sigma\Delta)^2(\Delta{J})^2/2}$, and $D_-(|\Delta{J}|)$
is reduced from $D_-(1)$ by a factor of
$e^{-(\Omega\sigma\Delta)^2(\Delta{J}-1)^2/2}$.}
\label{fig:pm}
\end{figure}

\end{document}